\pdfoutput=1
\documentclass[iop]{emulateapj}

\slugcomment{}

\shorttitle{Physical environment of IRAS 17008-4040 and IRAS 17009-4042 }
\shortauthors{L.~K. Dewangan et al.}
\begin{document}

\title{Investigating inner and large scale physical environments of IRAS 17008-4040 and IRAS 17009-4042 toward {\it l} = 345$\degr$.5, {\it b} = 0$\degr$.3}
\author{L.~K. Dewangan\altaffilmark{1}, T. Baug\altaffilmark{2}, D.~K. Ojha\altaffilmark{3}, and S.~K. Ghosh\altaffilmark{3}}
\email{Email: lokeshd@prl.res.in}
\altaffiltext{1}{Physical Research Laboratory, Navrangpura, Ahmedabad - 380 009, India.}
\altaffiltext{2}{Kavli Institute for Astronomy and Astrophysics, Peking 
University, 5 Yiheyuan Road, Haidian District, Beijing 100871, P. R. China.}
\altaffiltext{3}{Department of Astronomy and Astrophysics, Tata Institute of Fundamental Research, Homi Bhabha Road, Mumbai 400 005, India.}
\begin{abstract}
We present a multi-wavelength observational study of IRAS 17008-4040 and IRAS 17009-4042 to probe the star-formation (SF) mechanisms operational in both the sites. 
Each IRAS site is embedded within a massive ATLASGAL 870 $\mu$m clump ($\sim$2430--2900 M$_{\odot}$), and several parsec-scale filaments at 160 $\mu$m are radially directed toward these clumps (at T$_{d}$ $\sim$25--32 K). The analysis of the {\it Spitzer} and VVV photometric data depicts a group of infrared-excess sources toward both the clumps, suggesting the ongoing SF activities. 
In each IRAS site, high-resolution GMRT radio maps at 0.61 and 1.28 GHz confirm the presence of H\,{\sc ii} regions, which are powered by B-type stars. 
In the site IRAS 17008-4040, a previously known O-star candidate without an 
H\,{\sc ii} region is identified as an infrared counterpart of the 6.7 GHz methanol maser emission (i.e. IRcmme). Based on the VLT/NACO adaptive-optics L$^{\prime}$ image (resolution~$\sim$0\farcs1), the source IRcmme is resolved into two objects (i.e. IRcmme1 and IRcmme2) within a scale of 900~AU that are found to be associated with the ALMA core G345.50M. 
IRcmme1 is characterized as the main accreting HMPO candidate before the onset of an ultracompact H\,{\sc ii} region. In the site IRAS 17009-4042, the 1.28 GHz map has 
resolved two radio sources that were previously reported as a single 
radio peak. 
Altogether, in each IRAS site, the junction of the filaments (i.e. massive clump) is investigated with the cluster of infrared-excess sources and the ongoing massive SF.
These evidences are consistent with the ``hub-filament" systems as proposed by \citet{myers09}. 
\end{abstract}
\section{Introduction}
\label{sec:intro}
The birth of massive stars ($\gtrsim$ 8 M$_{\odot}$) is still an open research question in astrophysics although a significant progress has been 
made on both the theoretical and observational sides in recent years \citep[e.g.,][]{zinnecker07,tan14,motte17}. 
One of the competing theory among several others for the formation of massive star 
is ``high accretion of gas through the filaments", where massive stars are proposed 
to form at the junction of several such accreting filaments \citep[e.g.,][]{myers09,schneider12,yuan18,williams18}. 
In spite of their vast importance, study of the formation of the massive stars 
are elusive mainly because of their rarity, concealed pre-main-sequence phase 
and rather quick evolution compared to their low-mass counterparts. 
However, the 6.7 GHz methanol maser emission (mme) has been considered as one of the powerful tools for probing young massive stars \citep[e.g.,][]{walsh98,minier01,urquhart13}. Furthermore, the detection and absence of radio continuum emission toward the 6.7 GHz mme 
can enable us to study the earliest stages of massive star formation (MSF) prior to an ultracompact (UC) H\,{\sc ii} region, where one can examine the initial conditions of MSF \citep[e.g.,][]{tan14,dewangan15,motte17}. 
In the literature, we find such a potential site, IRAS 17008-4040 (G345.499+0.354) containing the 6.7 GHz mme, and the site is thought to host 
a genuine O-type protostellar object candidate \citep[e.g.,][]{morales09}.

Recently, \citet{lopez11} studied a giant molecular cloud (GMC) G345.5+1.0, which includes the IRAS point sources IRAS 17008-4040 and IRAS 17009-4042 (G345.490+0.311). 
Previously, an H\,{\sc ii} region was detected in each IRAS site using the Australia Telescope Compact Array (ATCA) radio continuum observations at 1.4 GHz and 2.5 GHz 
\citep{garay06}. The H\,{\sc ii} regions associated with IRAS 17008-4040 and IRAS 17009-4042 were reported to be excited by massive 
B0 and O9.5 stars, respectively \citep[e.g.,][]{garay06}. They adopted a distance of $\sim$2.0 kpc for both the IRAS sites. 
In the site IRAS 17008-4040, \citet{morales09} identified a bright and compact mid-infrared (MIR) source (i.e. IRAS 17008-4040~I) that 
was found toward the peak position of the dust continuum emission at 1.2 mm \citep[e.g.,][]{garay07,lopez11} and the 6.7 GHz mme \citep[e.g.,][]{walsh98}. 
However, the source IRAS 17008-4040~I was not associated with any radio 
continuum emission, and was seen with an extended 4.5 $\mu$m emission \citep[see Figure~4 in][]{morales09}. 
The extended 4.5 $\mu$m emission associated with IRAS 17008-4040~I was explained due to an outflow activity. \citet{morales09} also estimated its spectral type (i.e. O9.5) using the {\it Spitzer} MIR data (resolution $\sim$2$''$--6$''$) and the TIMMI2 data at 11.7--17.7 $\mu$m (resolution $\sim$1$''$), and characterized the source IRAS 17008-4040~I as a high mass protostellar object (HMPO) candidate. 
\citet{cesaroni17} examined the inner circumstellar environment (below 2000 AU) of G345.50+0.35 (i.e. the HMPO candidate IRAS 17008-4040~I) using the Atacama Large Millimeter/submillimeter Array (ALMA) observations with a resolution of $\sim$0$''$.2. 
They found two cores (i.e. G345.50M and G345.50S) in the continuum emission map at 218 GHz \citep[see Figure~2 in][]{cesaroni17}. 
Velocity gradients across these cores were also detected using the CH$_{3}$CN and $^{13}$CH$_{3}$CN lines \citep[see Figures~15 and~21 in][]{cesaroni17}. 
Based on the position-velocity analysis of these lines in the direction of both the cores, 
butterfly-shaped patterns were observed \citep[see Figures~22 and~23 in][]{cesaroni17}, and these results were interpreted as a signpost of Keplerian-like 
rotation in the cores. Hence, both the cores were reported as the best disc candidates by \citet{cesaroni17}. 
Most recently, \citet{urquhart18} cataloged the physical properties (i.e. velocities, distances, radii, and masses) of the 870 $\mu$m dust continuum clumps in the inner Galactic plane observed as a part of the APEX Telescope Large Area Survey of the Galaxy \citep[ATLASGAL; beam size $\sim$19$\farcs$2;][]{schuller09}. Using this publicly available catalog, we have also found several ATLASGAL clumps in the $\sim$0$\degr$.62 $\times$ 0$\degr$.62 area hosting both the IRAS sites 
(see Figure~\ref{sg1}a, and also Table~\ref{tab1} in this paper), which are found at a distance of 2.4 kpc and 
a molecular radial velocity (V$_{lsr}$) range of [$-$18, $-$15] km s$^{-1}$ \citep[e.g.,][]{urquhart18}. This distance estimate is almost in agreement with the previously adopted distance for the IRAS sources. Hence, in this paper, we 
have used the distance of 2.4 kpc for all the related analysis. 

Based on the literature survey, we find that the identification of filamentary features and their role in the star formation processes are yet to be studied in the IRAS sites. To understand the physical environment and star formation processes, a careful study of inner (below 2000 AU) and large (more than 20 pc) environments 
around both the IRAS sites is yet to be carried out. The physical processes concerning the existence of massive OB stars (including the HMPO candidate) and dust clumps are also not known. To investigate the physical processes in the sites IRAS 17008-4040 
and IRAS 17009-4042, we therefore present an extensive analysis of the multi-wavelength data sets (see Table~\ref{ftab1}), which also include the unpublished high-resolution near-infrared (NIR) images 
(resolutions $\sim$0$''$.1--0$''$.8) and radio continuum data (beam sizes $\sim$3$''$--8$''$). 
Furthermore, the adopted NIR data sets (resolution $\sim$0$''$.1) in this paper provide us an opportunity to examine the infrared morphology of the cores (i.e. G345.50M and G345.50S) observed by the ALMA data (resolution $\sim$0$''$.2).

Section~\ref{sec:obser} deals with the observational data sets and their analysis procedures. 
Section~\ref{sec:data} gives the outcomes of this paper. 
In Section~\ref{sec:disc}, we discuss the physical mechanisms operational in both the IRAS sites. 
Finally, the conclusions of this study are given in Section~\ref{sec:conc}.
\section{Data and analysis}
\label{sec:obser}
In this paper, we have employed the multi-wavelength data sets collected from various surveys, enabling us to probe the tens of parsecs to hundreds of AU environments of IRAS 17008-4040 and IRAS 17009-4042 (see Table~\ref{ftab1}). 
Some of the selected surveys (such as, 2MASS, VVV, GLIMPSE, Hi-GAL, ATLASGAL, and ThrUMMS) provide the processed data spanning from the NIR through radio wavelengths, which can be directly used for the scientific analysis.
The photometric magnitudes of point-like sources at VVV HK$_{s}$ and {\it Spitzer} 3.6--8.0 $\mu$m bands were extracted from the VVV DR2 \citep{minniti17} and the GLIMPSE-I Spring '07 highly reliable catalogs, respectively. 
Bright sources are saturated in the VVV survey \citep{minniti10,saito12,minniti17}. Hence, 2MASS photometric data were adopted for the bright sources. 
The photometric magnitudes of sources at {\it Spitzer} 24 $\mu$m were also obtained from the publicly available catalog \citep[e.g.,][]{gutermuth15}. In our selected target field, we also obtained the physical parameters of the ATLASGAL 870 $\mu$m dust continuum clumps from \citet{urquhart18}.

One can also note that some of the highlighted surveys (such as, GMRT and ESO VLT/NACO) give raw data, which are needed to be processed before performing any scientific analysis. 
In the following, we provide a brief description of the GMRT and the VLT/NACO data reduction procedures.
\subsection{Radio Continuum Observations}
Raw radio continuum data of an area hosting IRAS 17008-4040 and IRAS 17009-4042 
were obtained from the GMRT data archive in 0.61 and 1.28 GHz bands (Proposal Code: 11SKG01; PI: S.~K. Ghosh). 
The data were reduced using the Astronomical Image Processing System (AIPS) package following the standard procedures reported in \citet{mallick12,mallick13}. 
Bad data were flagged out from the UV data by multiple rounds of flagging using the {\sc tvflg} task of AIPS. After several 
rounds of `self-calibration', we finally obtained 0.61 and 1.28 GHz maps with the synthesized beams of 10$\farcs$1$\times$ 4$\farcs$6
and 5$\farcs$3$\times$1$\farcs$7, respectively. A correction arising due to different GMRT system temperatures for 
the two fields, viz., the science target field and the calibrator field, is required to be applied to the observed map. 
Especially, it becomes more vital for the sources located toward the Galactic plane. 
Because, the fluxes of such sources are generally calibrated using the flux calibrators that are located away from the Galactic plane. 
Thus, the background emission contributes more to the sources located toward the Galactic plane, and systematically increases the antenna temperature. 
It is particularly severe in low frequency bands (e.g., 0.61 GHz), where the contribution from the background emission is more. 
A detailed process of the system temperature correction can be found in \citet[][and references therein]{baug15}. 
We applied this correction in the 0.61 GHz map, before doing any scientific analysis. 
The final rms sensitivities of the 0.61 and 1.28 GHz maps are $\sim$0.3 and $\sim$0.4 mJy/beam, respectively.
\subsection{NIR adaptive-optics imaging data}
In the ESO-Science Archive Facility, the imaging observations of IRAS 17008-4040 in K$_{s}$- and L$^{\prime}$-bands are available (ESO proposal ID: 083.C-0582(A); PI: Jo\~{a}o Alves). 
These data sets were taken with the 8.2m VLT with NAOS-CONICA (NACO) adaptive-optics system \citep{lenzen03,rousset03}.
We processed these imaging data in this work. 
Following the same reduction processes outlined in \citet{dewangan15} and \citet{dewangan16b}, we produced the final processed VLT/NACO 
K$_{s}$ image (resolution $\sim$0\farcs2) and L$^{\prime}$ image (resolution $\sim$0\farcs1).
\section{Results}
\label{sec:data}
\subsection{Large scale physical environment}
\label{subsec:u0}
The observational study of a given star-forming region is often performed to infer its associated molecular cloud, dense clumps, and infrared-excess sources.  
\subsubsection{Molecular cloud and dust clumps}
\label{subsec:u1}
\begin{figure*}
\epsscale{0.77}
\plotone{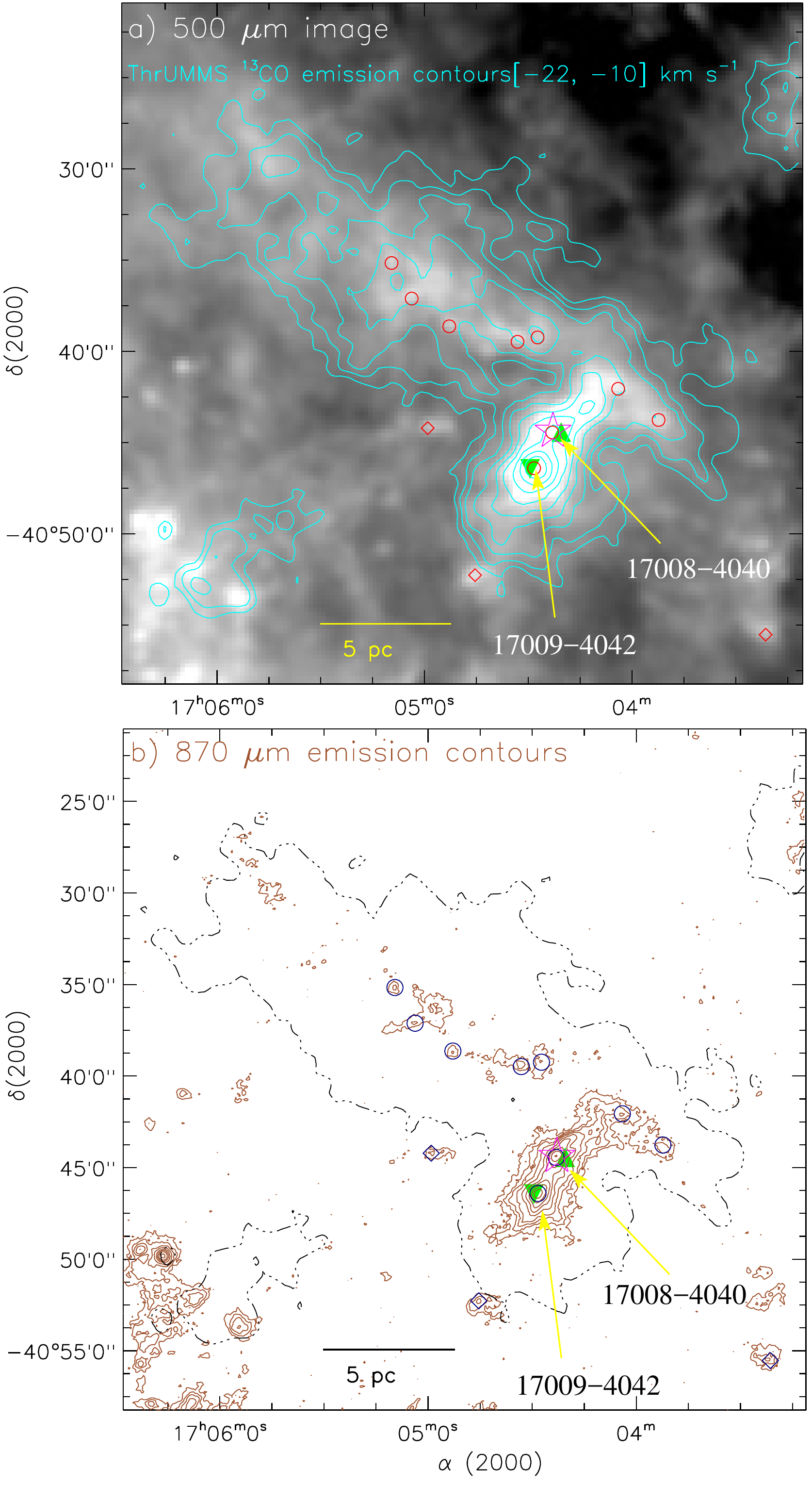}
\caption{a) Overlay of the molecular $^{13}$CO gas on the {\it Herschel} image at 500 $\mu$m (size $\sim$0$\degr$.62 $\times$ 0$\degr$.62; central coordinates: $\alpha_{2000}$ = 17$^{h}$04$^{m}$49.3$^{s}$, 
$\delta_{2000}$ = $-$40$\degr$39$\arcmin$43$\farcs8$). 
The $^{13}$CO emission is integrated over a velocity range of $-$22 to $-$10 km s$^{-1}$. 
The contours of $^{13}$CO (in cyan) are shown with the levels of 12.5, 16, 20, 26, 32, 40, 52, 64, and 75 K km s$^{-1}$. 
b) Overlay of the molecular $^{13}$CO gas on the ATLASGAL 870 $\mu$m dust continuum contours. 
The ATLASGAL dust continuum contours (in brown) are drawn with the levels of 0.13, 0.3, 0.5, 0.8, 1.4, 2.2, 3.4, 5.8, and 11 Jy/beam. The broken contour of $^{13}$CO (in black) is shown 
with a level of 12.5 K km s$^{-1}$. The positions of the observed ATLASGAL dust clumps at 870 $\mu$m \citep[from][]{urquhart18} are also marked in each figure (see circle and diamond symbols, and also Table~\ref{tab1}). 
Nine clumps highlighted with circles are traced in a velocity range of [-18, -15] km s$^{-1}$, while other three clumps marked with diamonds are traced in different V$_{lsr}$ values (i.e., $-$4.6, $-$26.2, and $-$23.8 km s$^{-1}$; see Table~\ref{tab1}). 
In both the panels, the positions of IRAS 17008-4040, IRAS 17009-4042, and 6.7 GHz mme 
are marked by triangle, upside down triangle, and star, respectively. 
The scale bar referring to 5 pc (at a distance of 2.4 kpc) is shown in both the panels.}
\label{sg1}
\end{figure*}

\begin{figure*}
\epsscale{1.18}
\plotone{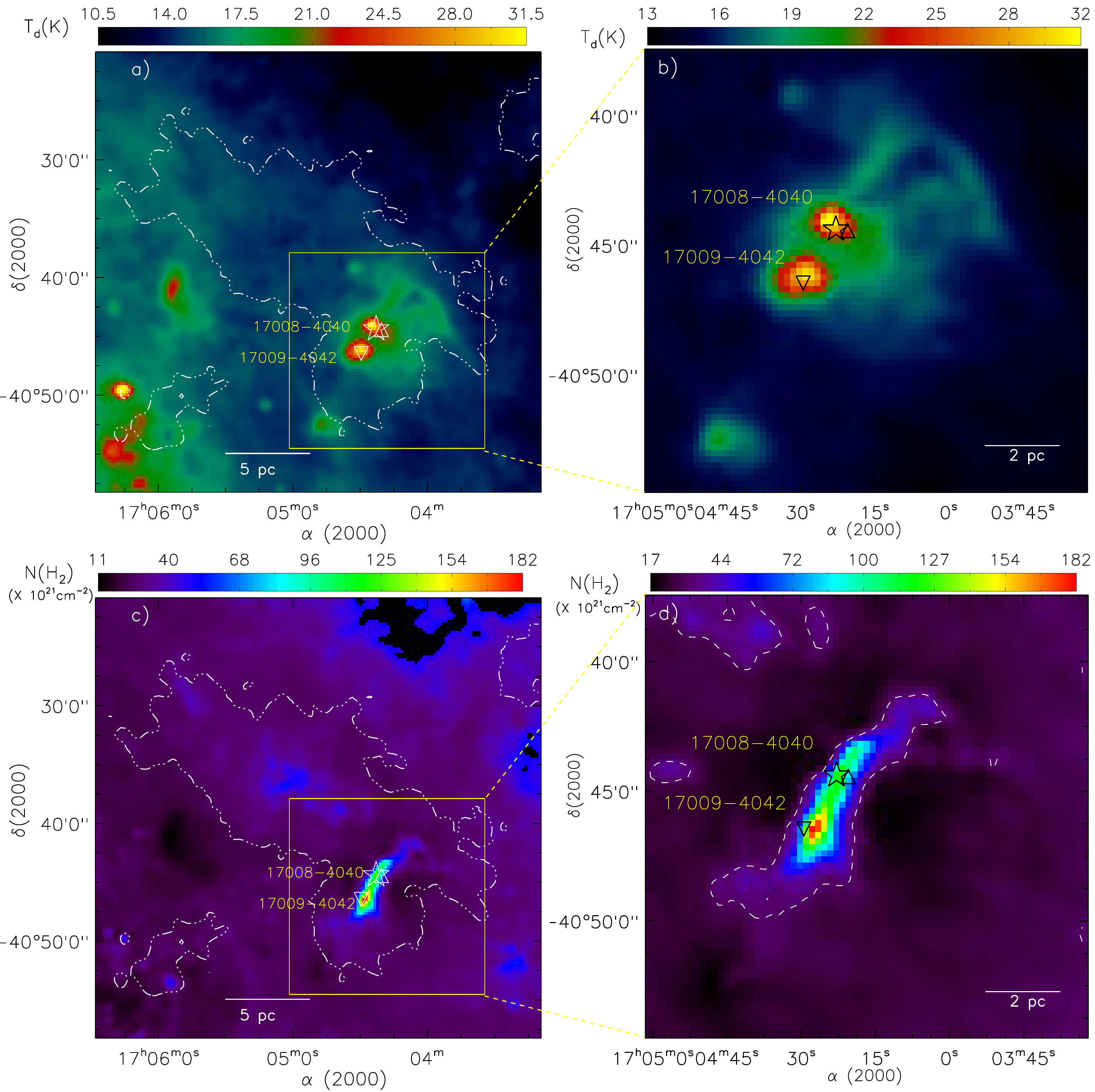}
\caption{a) {\it Herschel} temperature map (size $\sim$0$\degr$.62 $\times$ 0$\degr$.62; see Figure~\ref{sg1}a). 
b) A zoomed-in version of the {\it Herschel} temperature map (see a highlighted solid box in Figure~\ref{ysg2}a). c) {\it Herschel} column density map (size $\sim$0$\degr$.62 $\times$ 0$\degr$.62; see Figure~\ref{sg1}a).  
d) A zoomed-in version of the {\it Herschel} column density map (see a highlighted 
solid box in Figure~\ref{ysg2}c). A dashed $N(\mathrm H_2)$ contour (in white) with a level of 3.2 $\times$ 10$^{22}$ cm$^{-2}$ is also drawn in the panel ``d". In the panels ``a" and ``c", a dotted-dashed contour of $^{13}$CO (in white) is shown with a level of 12.5 K km s$^{-1}$. In all the panels, other symbols are the same as in Figure~\ref{sg1}.}
\label{ysg2}
\end{figure*}

In Figures~\ref{sg1} and~\ref{ysg2}, we examine the wide-field environment (i.e., $\sim$0$\degr$.62 $\times$ 0$\degr$.62) around IRAS 17008-4040 and IRAS 17009-4042.
Using the ThrUMMS $^{13}$CO line data, the molecular cloud associated with both the IRAS sites is studied in a velocity range of [$-$22, $-$10] km s$^{-1}$. 
The {\it Herschel} and ATLASGAL sub-mm dust continuum images are employed to study the embedded features and dust clumps in the molecular cloud.  
Figure~\ref{sg1}a shows the sub-mm image at 500 $\mu$m superimposed with the $^{13}$CO emission contours, indicating the boundary of the extended molecular cloud. 
The molecular cloud boundary and the ATLASGAL 870 $\mu$m dust continuum contours are shown in Figure~\ref{sg1}b. 
Using the integrated intensity map of $^{13}$CO at [$-$22, $-$10] km s$^{-1}$, the mass of the molecular 
cloud is determined to be $\sim$2.5 $\times$ 10$^{4}$ M$_{\odot}$. In the analysis, we employed an excitation temperature of 20 K, the ratio of gas to hydrogen by mass of about 1.36, and the abundance ratio (N(H$_{2}$)/N($^{13}$CO)) of 7 $\times$ 10$^{5}$ 
\citep[see][for more details]{yan16}. The positions of both the IRAS sources, the 6.7 GHz mme \citep[from][]{walsh98} and the ATLASGAL 870 $\mu$m dust continuum clumps \citep[from][]{urquhart18} are marked in Figures~\ref{sg1}a and~\ref{sg1}b. The sub-mm emission depicts the denser parts in the molecular cloud, and a majority of the sub-mm emission/dense material is 
concentrated in the direction of the sites IRAS 17008-4040 and IRAS 17009-4042. The sub-mm data reveal an elongated filamentary morphology 
containing both the IRAS sources. A total of 29 ATLASGAL clumps at 870 $\mu$m \citep{urquhart18} are found in the area of $\sim$0$\degr$.62 $\times$ 0$\degr$.62. Taking advantage of existing distance estimates of these clumps, we find 
only 12 clumps in the selected area at a distance of 2.4 kpc (see circles and diamonds in Figures~\ref{sg1}a and~\ref{sg1}b). In Table~\ref{tab1}, we have provided the physical parameters of these 12 clumps
(i.e. peak flux density, integrated flux density, V$_{lsr}$, distance, effective radius, dust temperature, and clump mass). 
Among the 12 clumps, we find nine dust clumps (i.e. c1--c9) traced in a velocity range of [$-$18, $-$15] km s$^{-1}$, which are highlighted by circles (see Figures~\ref{sg1}a and~\ref{sg1}b). 
We also marked remaining three dust clumps with diamonds (i.e. c10--c12), which have different V$_{lsr}$ values (i.e., $-$4.6, $-$26.2, 
and $-$23.8 km s$^{-1}$; see Table~\ref{tab1}) and are seen outside the molecular cloud boundary. 
Hence, these three clumps shown with diamonds do not appear to be part of the molecular cloud associated with the IRAS sites. 
Only nine ATLASGAL clumps (c1--c9) are distributed within the molecular cloud boundary, 
and the total mass of these nine clumps is $\sim$6605 M$_{\odot}$ (see Table~\ref{tab1}).  
A massive clump c1 (M$_{clump}$ = 2430 M$_{\odot}$) contains IRAS 17008-4040 and the 6.7 GHz mme, while IRAS 17009-4042 
is embedded in another massive clump c2 (M$_{clump}$ = 2900 M$_{\odot}$). The total mass of these two clumps (i.e. 5330 M$_{\odot}$) is about 21.3\% of the total molecular 
mass of the cloud. We have also computed virial mass (M$_{vir}$) and virial parameter (M$_{vir}$/$M_{clump}$) of these two massive clumps.
An expression of the virial mass of a clump of radius R$_{c}$ (in pc) and line width $\Delta V$ (in km s$^{-1}$) is given 
by M$_{vir}$ ($M_\odot$)\,=\,k\,R$_{c}$\,$\Delta V^2$ \citep{maclaren88}, where k\ (=\,126) is the geometrical parameter for a density profile $\rho$ $\propto$ 1/r$^2$. Using the $^{13}$CO line data, we have obtained the line widths toward the clumps c1 and c2 to 
be 1.7 and 2.38 km s$^{-1}$, respectively. 
Using the physical parameters of both the clumps (i.e., M$_{clump}$ and R$_{c}$; see Table~\ref{tab1}), 
the values of M$_{vir}$ for the clumps c1 and c2 are obtained to be $\sim$1107 and $\sim$1534 M$_{\odot}$. The analysis suggests that the virial parameters of both the clumps are less than 1.
It implies that both the clumps are unstable against gravitational collapse.

Figure~\ref{ysg2}a presents the {\it Herschel} temperature map (resolution $\sim$37$''$) in the direction of the molecular cloud associated 
with the IRAS sites. In the molecular cloud, an extended temperature structure is observed toward the elongated sub-mm morphology, 
and its zoomed-in view is shown in Figure~\ref{ysg2}b. Interestingly, the massive clumps c1 and c2 are depicted in a temperature range of about 25-32 K, and are surrounded by extended features at relatively low temperature of 19-22 K. 
Figure~\ref{ysg2}c shows the {\it Herschel} column density map (resolution $\sim$37$''$) of the molecular cloud associated with the IRAS sites, allowing us to analyse the column density distribution in the cloud. A zoomed-in view of the column density map is presented in Figure~\ref{ysg2}d. The {\it Herschel} column density map also enables us to infer the embedded structure of the cloud. Using the {\it Herschel} column density map, we can also obtain extinction \citep[$A_V=1.07 \times 10^{-21}~N(\mathrm H_2)$;][]{bohlin78} in the direction of the {\it Herschel} features/clumps. In the {\it Herschel} column density map, the elongated filamentary morphology is depicted with a $N(\mathrm H_2)$ contour level of 3.2 $\times$ 10$^{22}$ cm$^{-2}$ (or $A_V$ $\sim$34 mag), where both the IRAS sources are 
embedded (having peak $N(\mathrm H_2)$ = 1.82 $\times$ 10$^{23}$ cm$^{-2}$ (or $A_V$ $\sim$195 mag); see Figure~\ref{ysg2}d). 
The {\it Herschel} filamentary feature has a morphology very similar to that seen in the ATLASGAL 870 $\mu$m dust continuum map (see Figure~\ref{sg1}b). 
In order to generate the {\it Herschel} temperature and column density maps, we followed the steps given in \citet{mallick15}, and used the {\it Herschel} 160, 350, and 500 $\mu$m images in the analysis \citep[see also][]{dewangan18}. 
The image at 250 $\mu$m is saturated toward the IRAS positions, hence the data at 250 $\mu$m were excluded in the analysis. 
\subsubsection{Star formation in molecular cloud}
\label{ssubsec:u2}
\begin{figure*}
\epsscale{0.49}
\plotone{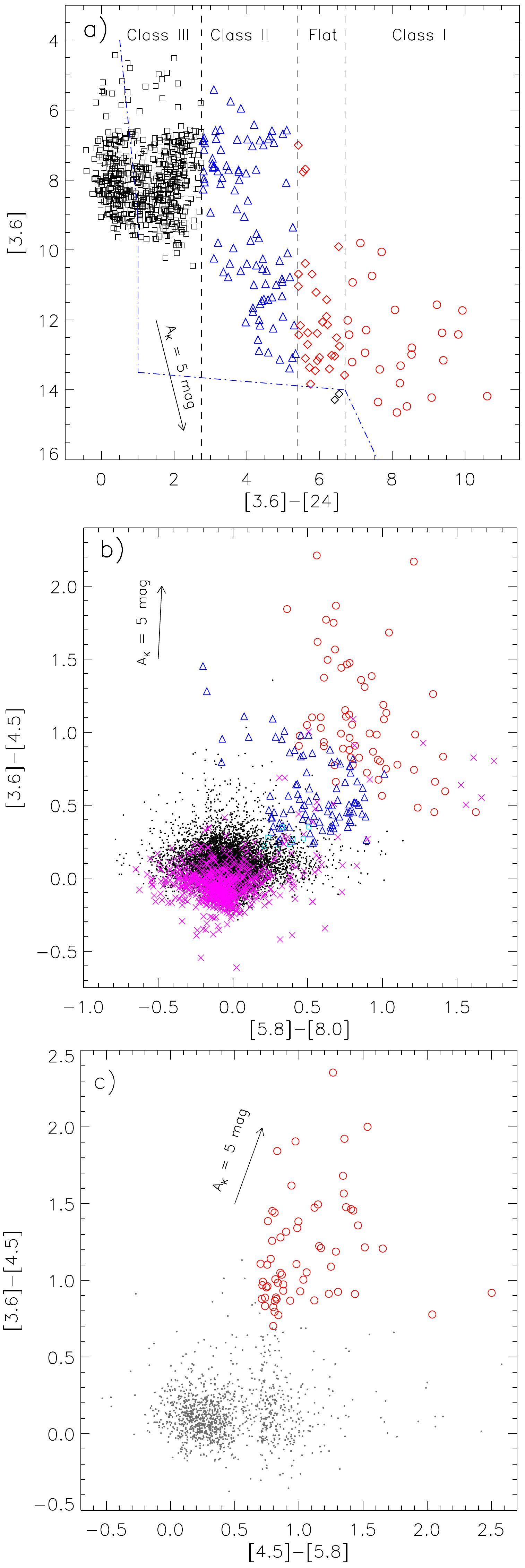}
\caption{a) {\it Spitzer} color-magnitude plot ([3.6] $-$ [24] vs [3.6]) of sources observed in our selected field (see Figure~\ref{sg1}a).
A broken curve (in blue) helps to separate YSOs against contaminated candidates (such as  galaxies and disk-less stars). 
With the help of dashed lines (in black), one can find YSOs belonging to different evolutionary stages. Symbols ``$\Diamond$'' and ``$\Box$'' refer to Flat-spectrum and Class~III sources, respectively. 
b) {\it Spitzer} color-color plot ([3.6]$-$[4.5] vs. [5.8]$-$[8.0]) of sources. 
The PAH-emission-contaminated apertures and Class~III sources are shown by ``$\times$'' (in magenta) and ``$\Box$'' (in cyan). 
c) {\it Spitzer} color-color plot ([4.5]$-$[5.8] vs [3.6]$-$[4.5]) of sources. 
An extinction vector \citep[from][]{flaherty07} is also drawn in each panel. 
In the panels ``b" and ``c", the dots refer to the stars with only photospheric emission. 
In each panel, Class~I and Class~II YSOs are represented by circles (in red) and triangles (in blue), respectively.}
\label{sg2}
\end{figure*}

To probe star formation activities in the molecular cloud associated with the IRAS sites, infrared-excess sources/young stellar objects (YSOs) are identified using the {\it Spitzer} color-magnitude and color-color plots. These plots also enable us to infer various contaminants (e.g. galaxies, disk-less stars, broad-line active galactic nuclei (AGNs), PAH-emitting galaxies, shocked emission blobs/knots, PAH-emission-contaminated apertures, and asymptotic giant branch (AGB) stars). Figures~\ref{sg2}a,~\ref{sg2}b, and~\ref{sg2}c show the {\it Spitzer} color-magnitude plot ([3.6]$-$[24]/[3.6]), color-color plot ([3.6]$-$[4.5] vs [5.8]$-$[8.0]), and color-color plot ([4.5]$-$[5.8] vs [3.6]$-$[4.5]), respectively. One can find more details of these plots in \citet{dewangan18}. 

First, we used the {\it Spitzer} color-magnitude plot ([3.6]$-$[24]/[3.6]) to select YSOs in our selected target field. In Figure~\ref{sg2}a, the boundaries of possible contaminants (i.e., galaxies and disk-less stars) and different stages of YSOs \citep[see][]{guieu10,rebull11} are marked. 
In the color-magnitude plot, Class~I, Flat-spectrum, and Class~II YSOs are shown by red circles, red diamonds, and blue triangles, respectively. 
Two Flat-spectrum sources are excluded from our selected YSO list, and are seen in the contaminants zone (see black diamonds in Figure~\ref{sg2}a). 
In our selected YSO catalog, we also applied a condition (i.e. [4.5] $>$ 7.8 mag and [8.0]-[24.0] $<$ 2.5 mag) to know the possible AGB contaminants \citep[e.g.,][]{robitaille08}.
This analysis yields 19 possible AGB contaminants, which are not considered in the final catalog. 
Hence, we find a total of 130 YSOs (25 Class~I, 29 Flat-spectrum, and 76 Class~II) in our selected YSO catalog. 

In Figure~\ref{sg2}b, the selected Class~I and Class~II YSOs are marked by red circles and blue triangles, respectively. To select YSOs and various contaminants in the color-color plot, we followed the steps given in \citet{gutermuth09} and \citet{lada06} \citep[see also][for more details]{dewangan18}. 
Using the {\it Spitzer} 3.6, 4.5, 5.8, and 8.0 $\mu$m photometric data, the color-color plot yields a total of 71 YSOs (30 Class~I and 41 Class~II). 
These additional YSOs are not overlapped with the YSOs identified using the {\it Spitzer} color-magnitude plot ([3.6]$-$[24]/[3.6]).

In Figure~\ref{sg2}c, the selected Class~I YSOs are marked by red circles in the plot. 
These YSOs are identified with the infrared color conditions (i.e. [4.5]$-$[5.8] $\ge$ 0.7 mag and [3.6]$-$[4.5] $\ge$ 0.7 mag), which are taken from \citet{hartmann05} and \citet{getman07}. Using the first three {\it Spitzer}-GLIMPSE bands, the color-color plot yields a total of 63 Class~I YSOs in our selected region. Furthermore, these additional YSOs are also not common with the YSOs identified using the color-magnitude plot ([3.6]$-$[24]/[3.6]) and the color-color plot ([3.6]$-$[4.5] vs [5.8]$-$[8.0]). 

\begin{figure*}
\epsscale{0.82}
\plotone{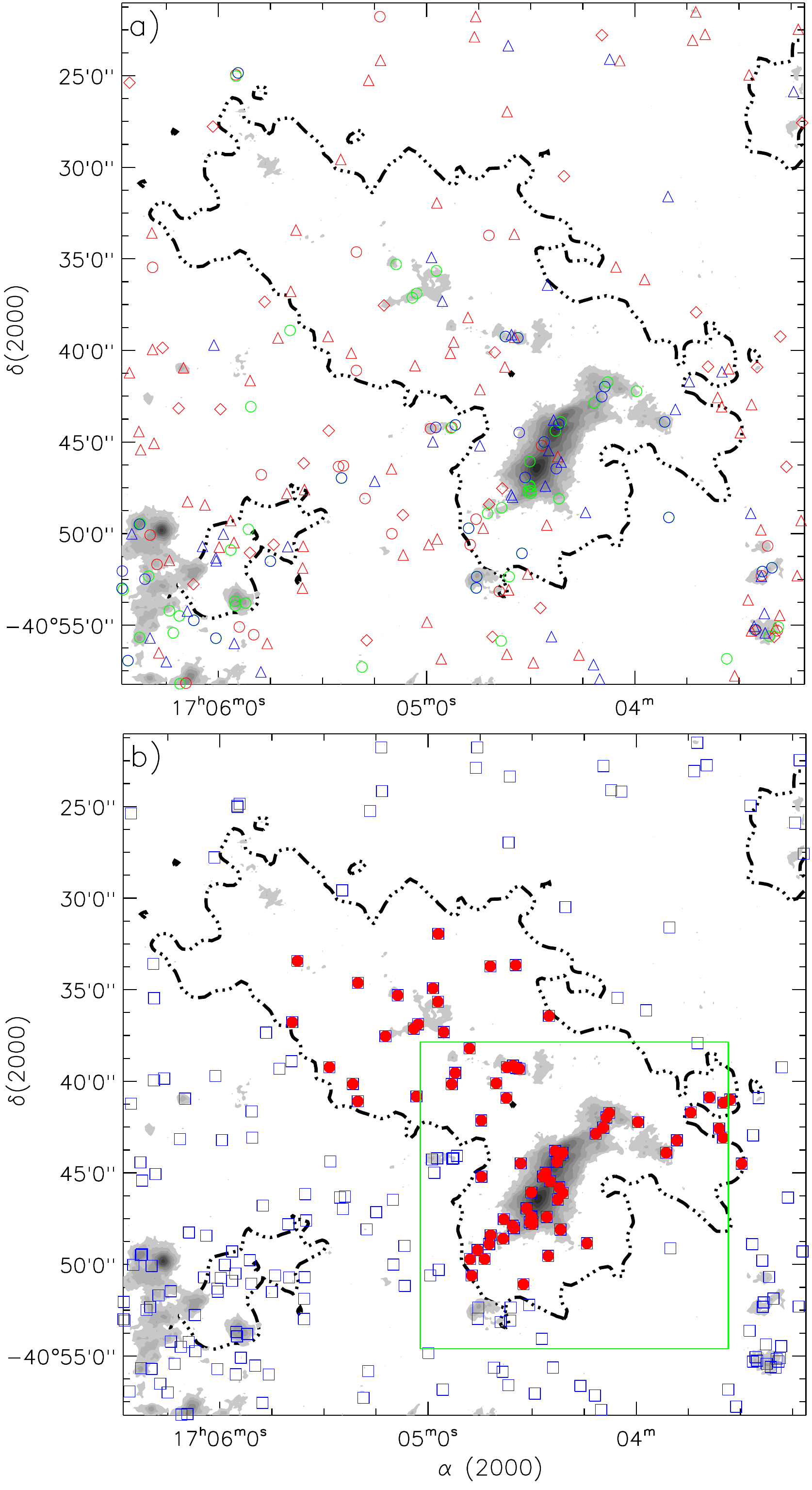}
\caption{Spatial distribution of YSOs observed in our selected field (see Figure~\ref{sg1}a). 
a) Overlay of the selected YSOs (Class~I (circles), Flat-spectrum (diamond), and Class~II (triangles)) on the $^{13}$CO emission and the ATLASGAL 870 $\mu$m 
dust continuum contour map (see also Figure~\ref{sg1}). 
The YSOs (in red) are selected using the color-magnitude ([3.6] $-$ [24] vs [3.6]; see Figure~\ref{sg2}a), while the YSOs (in blue) identified using the color-color plot ([3.6]$-$[4.5] vs. [5.8]$-$[8.0]; see Figure~\ref{sg2}b). The YSOs (in green) are selected using 
the color-color plot ([4.5]$-$[5.8] vs [3.6]$-$[4.5]; see Figure~\ref{sg2}c).
b) Filled squares (in red) indicate the YSOs distributed within the molecular cloud boundary, while the YSOs outside the molecular cloud boundary are marked by open squares (in blue).}
\label{ysg3}
\end{figure*}

In the final YSO catalog, we find a total of 264 YSOs in our selected region, which are overlaid on the $^{13}$CO and 870 $\mu$m dust continuum contour maps (see Figure~\ref{ysg3}a). 
In Figure~\ref{ysg3}b, we have shown the filled and open squares to 
highlight the selected YSOs located inside and outside the molecular cloud, respectively. 
A total of 78 YSOs are spatially found inside the molecular cloud boundary (see Figure~\ref{ysg3}b), which are unlikely to be contaminated by field stars along the line of sight.  
These embedded YSOs are mainly found toward the denser regions traced by the sub-mm dust 
emission within the molecular cloud, indicating the areas of the ongoing star formation 
in the molecular cloud. One can also notice that a majority of these YSOs are distributed toward the 
major axis of the elongated filamentary morphology containing the massive clumps c1 and c2 (see Figure~\ref{ysg3}b).
\subsubsection{Molecular condensations and position-velocity plots of CO}
\label{ssubsec:u3}
\begin{figure*}
\epsscale{1}
\plotone{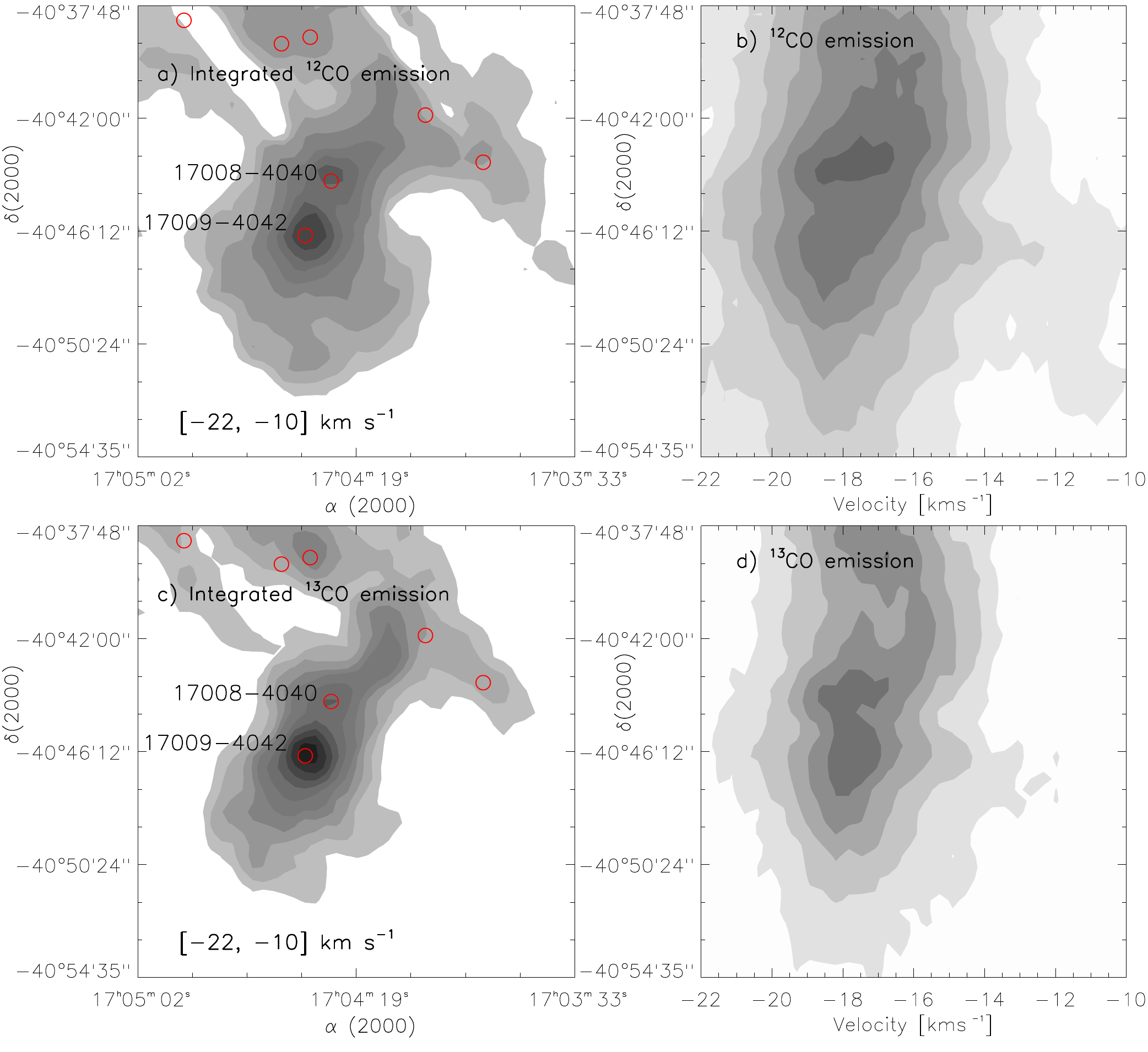}
\caption{Distribution of $^{12}$CO and $^{13}$CO emission toward the region around IRAS 17008-4040 
and IRAS 17009-4042 (see the dotted-dashed box in Figure~\ref{ysg3}b; 
size$\sim$16$\farcm8$ $\times$ 16$\farcm8$; 
centered at $\alpha_{2000}$ = 17$^{h}$04$^{m}$17.7$^{s}$, $\delta_{2000}$ = $-$40$\degr$46$\arcmin$14$\farcs4$).
The contour maps of integrated $^{12}$CO emission (see top left panel ``a") and $^{13}$CO (see bottom 
left panel ``c") emission in a velocity range of $-$22 to $-$10 km s$^{-1}$. Declination-velocity maps of $^{12}$CO 
(see top right panel ``b") and $^{13}$CO (see bottom right panel ``d"). 
In the panel ``a", the contour levels of $^{12}$CO are 30, 35, 40, 50, 60, 70, 80, and 90\% of the peak value (i.e. 190.86 K km s$^{-1}$).
In the panel ``c", the contour levels of $^{13}$CO are 20, 25, 30, 35, 40, 50, 60, 70, 80, and 90\% of the peak value (i.e. 92.75 K km s$^{-1}$).
The ATLASGAL dust continuum clumps at 870 $\mu$m \citep[from][]{urquhart18} are also overlaid on each molecular intensity map (see panels ``a" and ``c").}
\label{sg3}
\end{figure*}

In Figure~\ref{ysg3}b, a solid box highlights the area, where the $^{12}$CO and $^{13}$CO emissions are 
prominent. In the direction of this selected area, Figures~\ref{sg3}a and~\ref{sg3}c present the $^{12}$CO and $^{13}$CO intensity maps, respectively. The molecular condensations are seen toward the locations 
of the IRAS sources. 
Figures~\ref{sg3}b and~\ref{sg3}d show the declination-velocity plots of $^{12}$CO and $^{13}$CO, respectively. 
In the velocity space, a noticeable molecular velocity spread is seen in the direction of the elongated morphology. 

\begin{figure*}
\epsscale{0.82}
\plotone{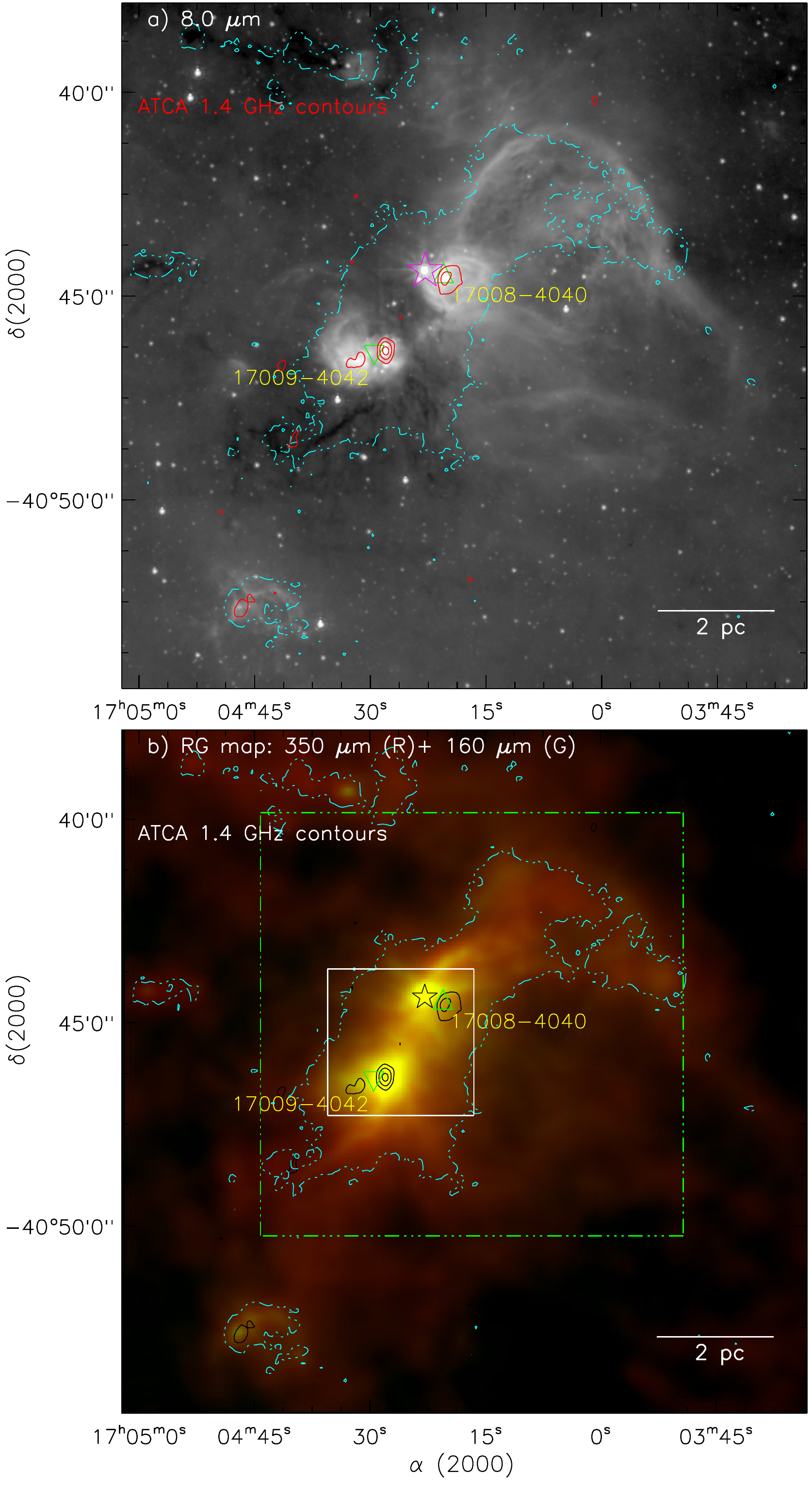}
\caption{a) Overlay of the ATLASGAL 870 $\mu$m dust continuum contour (in cyan) and the ATCA 1.4 GHz continuum contours \citep[in red; angular resolution $\sim$10$''$.1 $\times$ 6$''$.1;][]{garay06} on the {\it Spitzer} 8.0 $\mu$m image (see the dotted-dashed box in Figure~\ref{sg2}b). 
b) Overlay of the ATLASGAL 870 $\mu$m dust continuum contour (in cyan) and the ATCA 1.4 GHz continuum contours (in black) on a two color-composite map. 
The color-composite map is the result of the combination of two bands: 350 $\mu$m (red) and 160 $\mu$m (green). In each panel, the ATLASGAL 870 $\mu$m dust continuum contour is drawn with a level of 0.13 Jy/beam, and the ATCA 1.4 GHz continuum contours are shown with the levels of 3.96, 39.59, and 237.53 mJy/beam. Other symbols are the same as in Figure~\ref{sg1}.}
\label{sg4}
\end{figure*}

In Figure~\ref{sg4}a, we have shown the {\it Spitzer} 8.0 $\mu$m image overlaid with the ATLASGAL 870 $\mu$m continuum contour 
and the ATCA 1.4 GHz radio continuum emission. 
The radio continuum emission traces the H\,{\sc ii} 
regions toward both the IRAS sources located well within the elongated and extended sub-mm morphology. 
In each IRAS position, an extended 8.0 $\mu$m structure containing the H\,{\sc ii} region is seen, which has also been reported earlier \citep[e.g.,][]{morales09,lopez11}. 
In this work, the proposed HMPO candidate IRAS 17008-4040~I is considered as an infrared counterpart of the 6.7 GHz mme (hereafter IRcmme), 
and is not associated with the ionized emission. 
The source is detected in the 2MASS K$_{s}$ and all {\it Spitzer}-GLIMPSE bands. 
However, the source is saturated in the {\it Spitzer} 8.0 $\mu$m image. 
Using the {\it Spitzer} photometric data at 3.6--5.8 $\mu$m, 
the source IRcmme is identified as a protostar. 
Figure~\ref{sg4}b shows a two color-composite map made using the {\it Herschel} 350 $\mu$m (in red) and 
{\it Herschel} 160 $\mu$m (in green) images. The color-composite map hints the presence of several {\it Herschel} filaments within the elongated morphology. 
\subsection{Hub-filament systems}
\label{subsec:u2}
\begin{figure*}
\epsscale{0.55}
\plotone{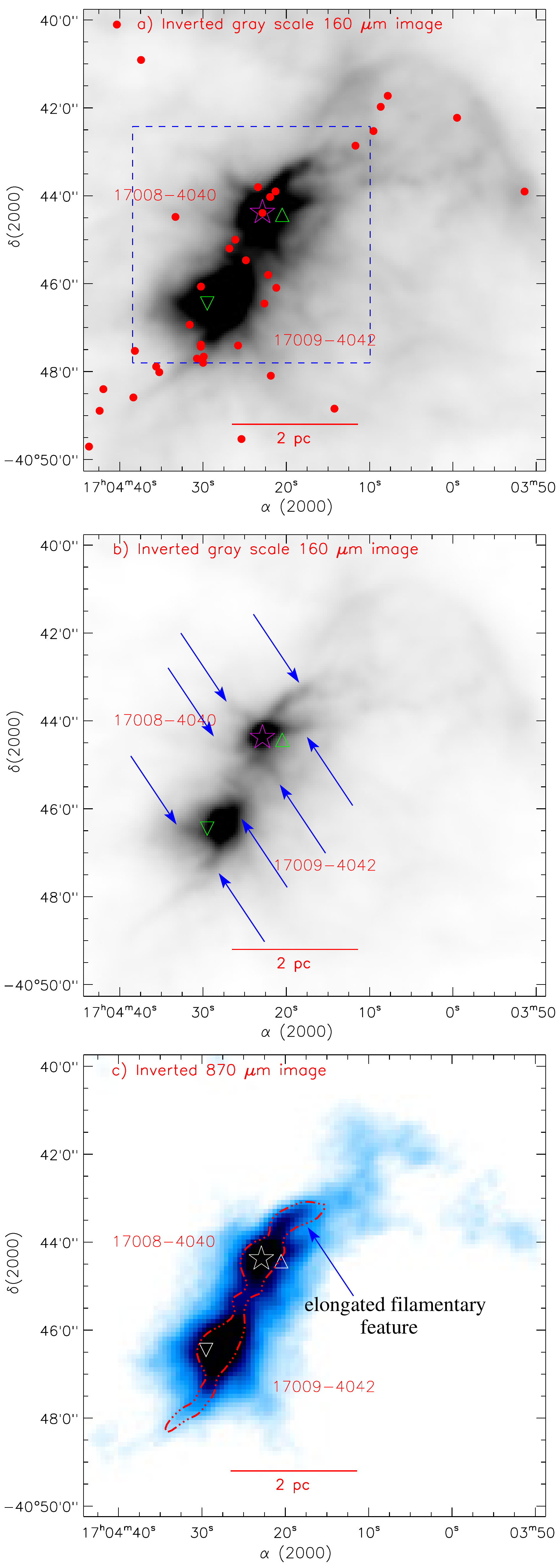}
\caption{a) Overlay of the selected YSOs (red filled circles) 
on an inverted gray scale {\it Herschel} 160 $\mu$m image (see the dotted-dashed box in Figure~\ref{sg4}b). These YSOs are distributed within the molecular cloud boundary, and 
are taken from Figure~\ref{ysg3}b. 
b) The panel displays an inverted gray scale {\it Herschel} 160 $\mu$m image (resolution $\sim$12$''$). 
The embedded faint filament-like features are also highlighted by blue arrows. 
c) The panel displays an inverted ATLASGAL 870 $\mu$m image (beam size $\sim$19$''$.2). 
An elongated filamentary feature is highlighted by a broken contour (in red). 
In each panel, other symbols are the same as in Figure~\ref{sg1}.}
\label{sg5}
\end{figure*}

In Figure~\ref{sg5}a, we have shown an inverted gray scale {\it Herschel} 160 $\mu$m image 
overlaid with the selected YSOs. 
Several faint filament-like features are prominently seen in the {\it Herschel} 160 $\mu$m image (see Figure~\ref{sg5}b). An inverted 870 $\mu$m image (in blue) is shown in Figure~\ref{sg5}c. 
An elongated filamentary feature is highlighted by a broken contour in Figure~\ref{sg5}c. 
Interestingly, due to much better resolution of the {\it Herschel} 160 $\mu$m image compared to the ATLASGAL 870 $\mu$m image, at least three faint {\it Herschel} filaments/fibres appear to be radially directed to the ATLASGAL clumps c1 and c2 
(see Figure~\ref{sg5}b). These results indicate the presence of a probable ``hub-filament" system toward 
IRAS 17008-4040 and IRAS 17009-4042 (see blue arrows in Figure~\ref{sg5}b). 
One can find the implication of these observed results in Section~\ref{subsec:u5}. 
\subsection{Ionized clumps and clustering of sources}
\label{subsec:u3}
\begin{figure*}
\epsscale{1.15}
\plotone{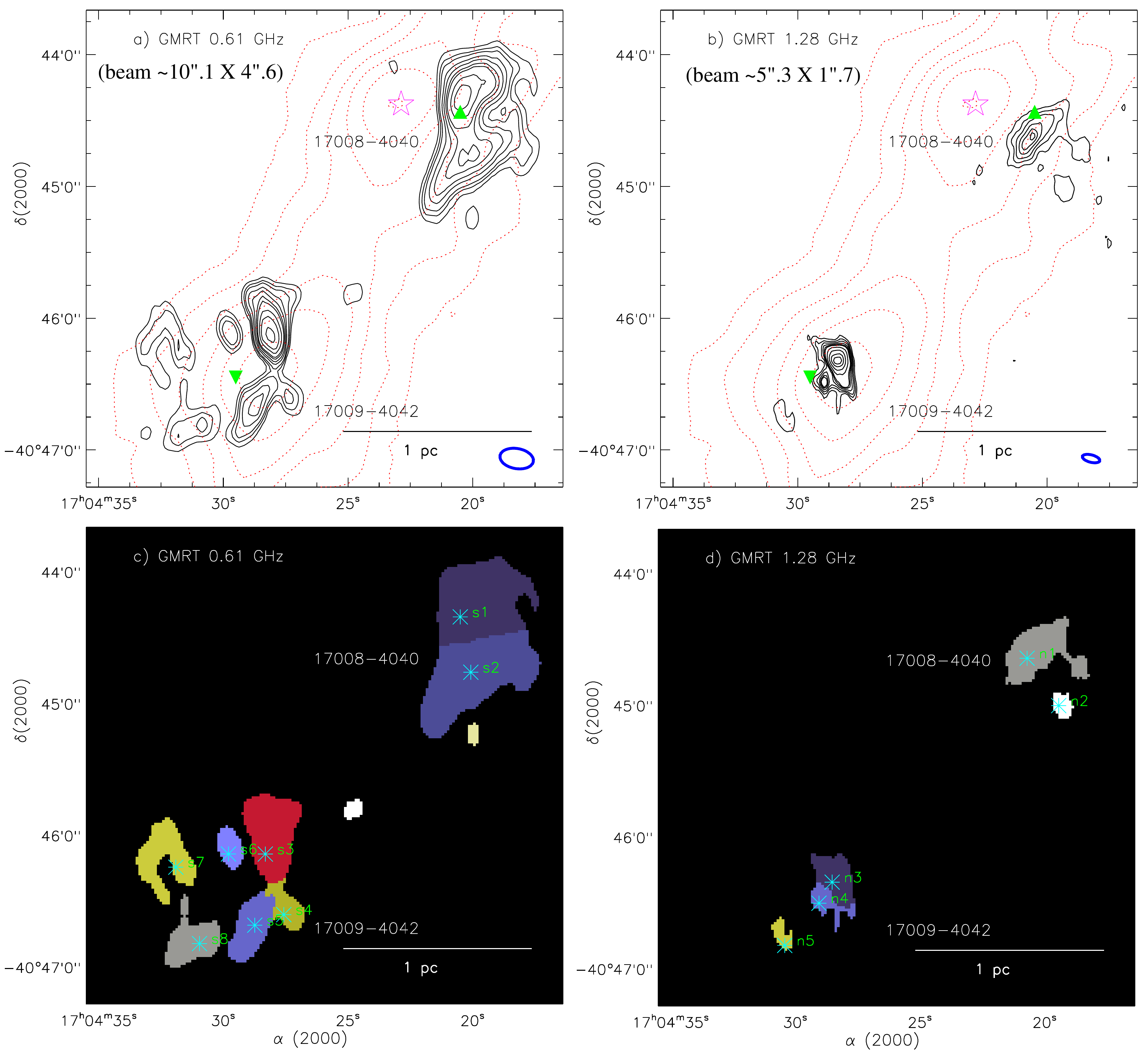}
\caption{High-resolution radio continuum maps at two GMRT frequencies toward IRAS 17008-4040 and IRAS 17009-4044 (see the solid box in Figure~\ref{sg4}b). 
a) GMRT 0.61 GHz continuum contours (in black; beam size $\sim$10$''$.1 $\times$ 4$''$.6; sensitivity $\sim$0.3 mJy/beam) are shown with the levels of 1.48, 2.5, 3.5, 5.5, 7.5, 10, 12, 17, 22, and 28 mJy/beam. 
b) GMRT 1.28 GHz continuum contours (in black; beam size $\sim$5$''$.3 $\times$ 1$''$.7; sensitivity $\sim$0.4 mJy/beam) are shown with the levels of 3.5, 6.5, 9.5, 13, 16, 18, 25, 35, 45, and 52 mJy/beam. 
c--d) The boundary of each identified clump in both the GMRT radio maps is highlighted along with its corresponding clump ID (see Table~\ref{stab1} and also Figures~\ref{sg77}a and~\ref{sg77}b). In the panels ``a" and ``b", the ATLASGAL 870 $\mu$m dust continuum contours (in red) are displayed with 
the levels of 0.8, 1.4, 2.2, 3.4, 5.8, and 11 Jy/beam, and a star symbol indicates the position of the 6.7 GHz mme. 
Ellipses (in blue) represent the beam sizes of radio continuum data in the panels ``a" and ``b". In all the panels, other symbols are the same as in Figure~\ref{sg1}.}
\label{sg77}
\end{figure*}

To examine the H\,{\sc ii} regions/ionized clumps in both the IRAS sites, we present high-resolution GMRT radio continuum maps at 0.61  GHz (beam size $\sim$10$''$.1 $\times$ 4$''$.6; sensitivity $\sim$0.3 mJy/beam) and 1.28 GHz (beam size $\sim$5$''$.3 $\times$ 1$''$.7; sensitivity $\sim$0.4 mJy/beam) in Figures~\ref{sg77}a and~\ref{sg77}b, respectively. 
The GMRT radio map at 1.28 GHz has higher spatial resolution compared to the map at 0.61 GHz. Hence, the map at 1.28 GHz provides more insights into the individual clumps. 
However, the GMRT 0.61 GHz radio map reveals several extended ionized features compared to the map at 1.28 GHz. In both the figures, we have also shown the ATLASGAL 870 $\mu$m dust emission contours and the position of the 6.7 GHz mme. In the direction of IRAS 17008-4040, there is no ionized clump seen toward the peak positions of the ATLASGAL 870 $\mu$m dust emission and the 6.7 GHz mme. However, the ionized emission is observed toward the position of IRAS 17008-4040, and is about 29$''$ away from the 6.7 GHz mme.
On the other hand, the ionized clump is observed toward the peak position of the ATLASGAL 870 $\mu$m dust emission in the direction of IRAS 17009-4042. 

\begin{figure*}
\epsscale{1.15}
\plotone{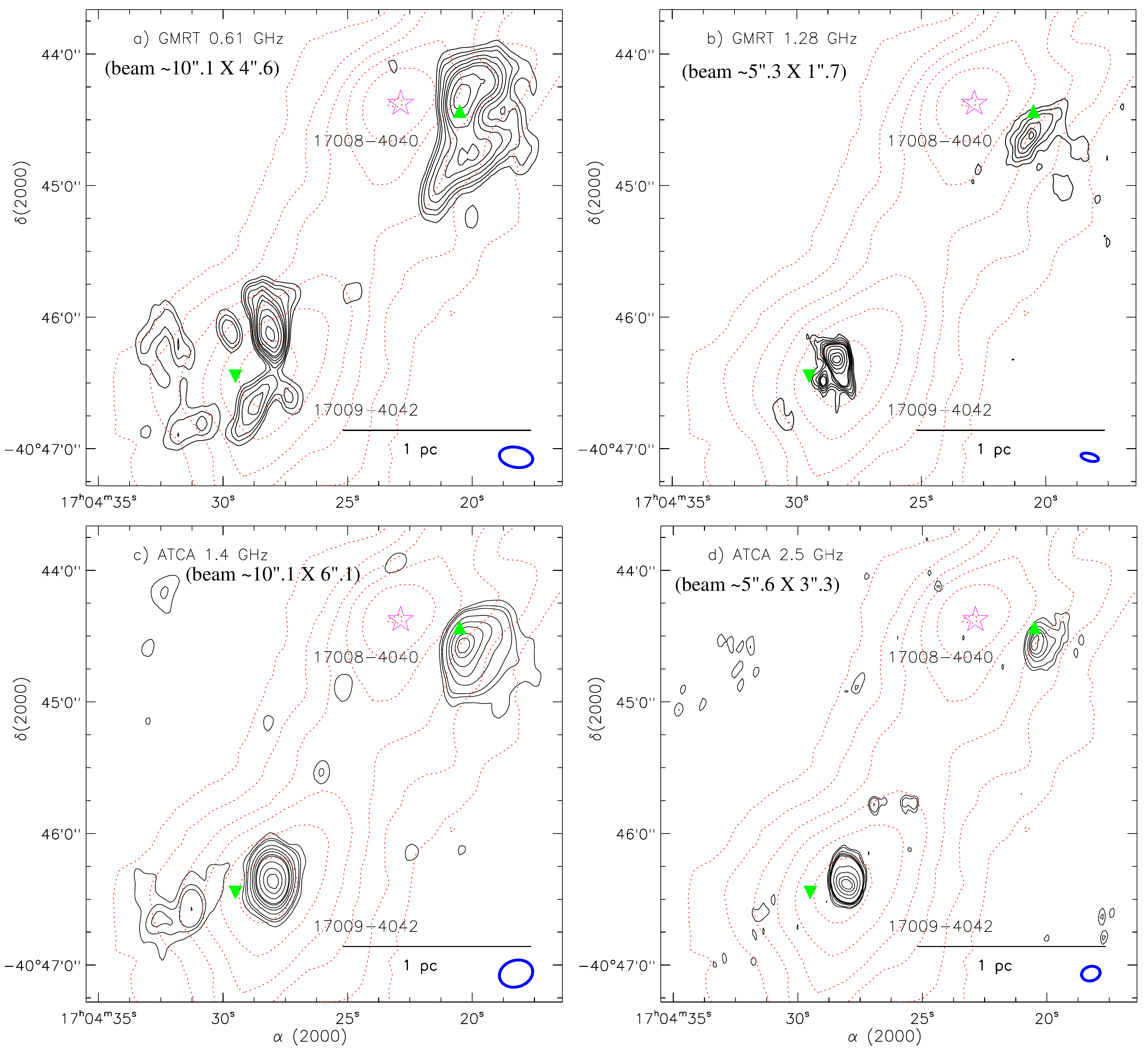}
\caption{Radio continuum emission contours at different frequencies toward IRAS 17008-4040 and IRAS 17009-4044 (see the solid box in Figure~\ref{sg4}b). 
a) GMRT 0.61 GHz continuum contours (see Figure~\ref{sg77}a). 
b) GMRT 1.28 GHz continuum contours (see Figure~\ref{sg77}b). 
c) ATCA 1.4 GHz continuum contours (in black; beam size $\sim$10$''$.1 $\times$ 6$''$.1) are shown with the levels of 2, 4, 6, 15, 28, 43, 66, 80, 150, 250, and 340 mJy/beam \citep[see also][]{garay06}. 
d) ATCA 2.5 GHz continuum contours (in black; beam size $\sim$5$''$.6 $\times$ 3$''$.3) are shown with the levels of 1.6, 2.6, 6.2, 9.5, 12, 15, 60, 140, 250, and 330 mJy/beam \citep[see also][]{garay06}. 
In each panel, the ATLASGAL 870 $\mu$m dust continuum contours (in red) are displayed with 
the levels of 0.8, 1.4, 2.2, 3.4, 5.8, and 11 Jy/beam. In all the panels, ellipses (in blue) represent the beam sizes of radio continuum data. In each panel, a star symbol indicates the position of the 6.7 GHz mme. In all the panels, other symbols are the same as in Figure~\ref{sg1}.}
\label{ttsg77}
\end{figure*}

We have identified eight ionized clumps (s1--s8) in the GMRT 0.61 GHz radio map (see Figure~\ref{sg77}c), while five radio sources (n1--n5) are identified in the 1.28 GHz map 
(see Figure~\ref{sg77}d). Following the procedure reported in \citet{dewangan17a}, we estimated the 
Lyman continuum photons \citep[see also][for equation]{matsakis76} and spectral type of each radio source in the GMRT maps.  
In the analysis, we adopted a distance of 2.4 kpc, an electron temperature of 10$^{4}$ K, and the models of \citet{panagia73}. Accordingly, we found that all the 
ionized clumps are powered by massive B-type stars. 
We have tabulated the derived physical properties of the ionized clumps (i.e. deconvolved effective radius of the ionized clump ($R_\mathrm{HII}$), total flux (S${_\nu}$), 
Lyman continuum photons (log$N_\mathrm{uv}$), and radio spectral type) in Table~\ref{stab1}. 
In Figure~\ref{ttsg77}, the GMRT maps are compared with the ATCA 1.4 and 2.5 GHz radio continuum maps. 
The GMRT map at 1.28 GHz shows almost similar radio morphology to those detected in the ATCA 1.4 and 2.5 GHz radio continuum maps toward the IRAS sites. However, the GMRT map at 1.28 GHz resolves the previously detected a single IRAS 17009-4042 H\,{\sc ii} region into two ionized clumps (see n3 and n4 in Figure~\ref{sg77}d). 
Furthermore, the radio morphology seen in the low-frequency map at 0.61 GHz appears 
little different from that of other radio continuum maps. 
A reasonable explanation of the observed feature in the 0.61 GHz map could be that radio emission at lower frequencies would be more
sensitive to more diffuse ionized gas \citep[see][and references therein]{yang18}.  
Another reason could be the short-spacing problem of interferometric observations \citep[see][]{thompson01,stanimirovic02}. 
Due to these reasons we see the different radio continuum structure in the 0.61 GHz map compared to other ATCA and GMRT maps.

\begin{figure*}
\epsscale{0.69}
\plotone{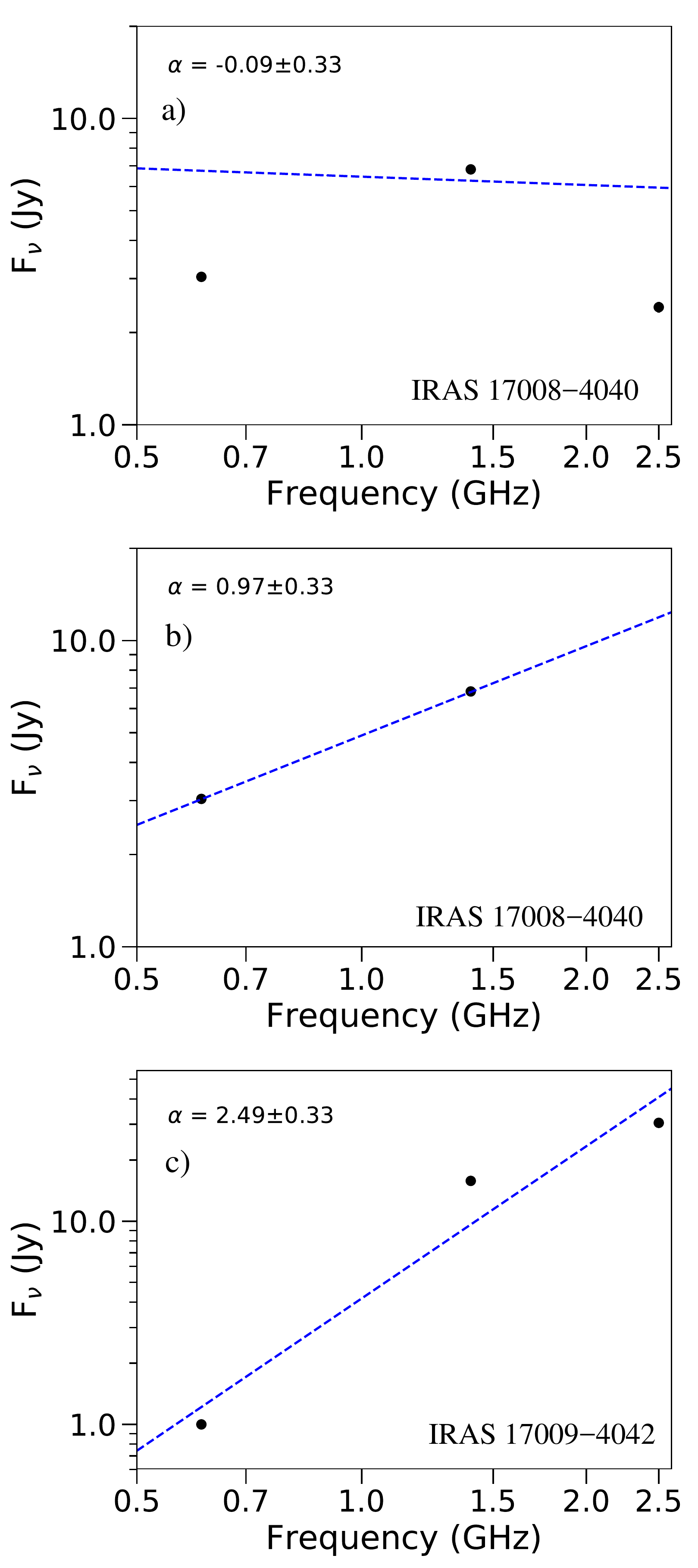}
\caption{a) Radio spectral index plot of the radio clump associated with IRAS 17008-4040. 
Filled circles (in black) are the flux densities at 0.61, 1.4, and 2.5 GHz. 
b) Same as Figure~\ref{xy10}a, but only two flux densities at 0.61 and 1.4 GHz are considered. 
c) Radio spectral index plot of the radio clump associated with IRAS 17009-4042.
Filled circles (in black) are the flux densities at 0.61, 1.4, and 2.5 GHz.}
\label{xy10}
\end{figure*}

The knowledge of a radio spectral index of a given radio source is very useful to acquire the information of 
ongoing radio emission process in the source. The radio spectral index ($\alpha$) is defined as F$_\nu$ $\propto$ $\nu^{\alpha}$, where $\nu$ is the frequency of observation, and F$_\nu$ is the corresponding observed flux density. 
As seen in Figure~\ref{ttsg77}, the radio continuum observations at four frequencies are detected toward both the IRAS sources. 
To determine the spectral indices of the ionized clumps associated with IRAS 17008-4040 and IRAS 17009-4042, firstly, 
all the radio continuum maps are convolved to the same (lowest) resolution of 11$''$ $\times$ 6$''$.5. 
Then, we use the {\sc jmfit} task of AIPS on all the convolved radio continuum maps to estimate the flux densities 
and sizes of the observed ionized clumps. However, we find that the flux densities of radio clumps in 
the GMRT 1.28 GHz map is relatively higher than that of the ACTA 1.4 GHz map. Hence, we have preferred the flux densities at 1.4 GHz in the spectral index analysis. 
In Figure~\ref{xy10}a, the radio spectral index plot of the radio clump associated with IRAS 17008-4040 is presented using three flux densities, however the fit is not very good. 
Figure~\ref{xy10}b shows the radio spectral index plot of the IRAS 17008-4040 clump using only two flux densities. 
We find the spectral index for the IRAS 17008-4040 clump to be $<$ 1.0 (having a range of $-$0.09 to 0.97). 
Such flat spectral index indicates the presence of non-thermal contribution in addition to the free-free emission in the IRAS 17008-4040 clump. In Figure~\ref{xy10}c, the radio spectral index plot of the radio clump associated with IRAS 17009-4042 is shown 
using three flux densities. However, the fit gives a spectral index of 2.49$\pm$0.33. 
In the case of the IRAS 17009-4042 clump, the observed spectral index implies the thermal free-free emission originated in an optically thick medium.

\begin{figure*}
\epsscale{0.52}
\plotone{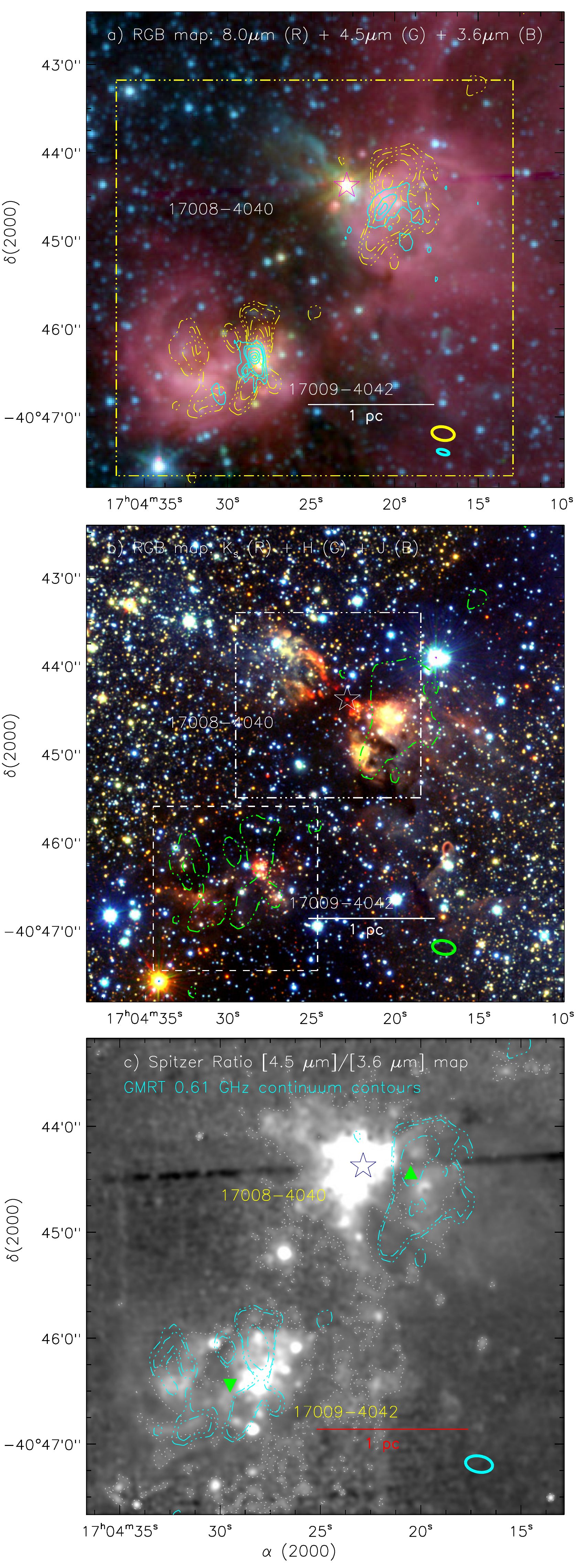}
\caption{a) Overlay of the GMRT 0.61 GHz continuum contours (in orange) 
and 1.28 GHz continuum contours (in cyan) on a three color-composite map (see the dashed box in Figure~\ref{sg5}a and also Figure~\ref{sg77}). 
The color-composite map is the result of the combination of three bands: 8.0 $\mu$m (red), 4.5 $\mu$m (green), and 3.6 $\mu$m (blue). b) Overlay of the GMRT 0.61 GHz continuum contour (in green; see also Figure~\ref{sg77}) on a three color-composite map (see the dashed box in Figure~\ref{sg5}a). 
The color-composite map is the result of the combination of three VVV bands: K$_{s}$ (red), H (green), and J $\mu$m (blue). The dotted-dashed box (in white) encompasses the area shown in Figure~\ref{sg8}a, while the dashed box (in white) refers the area shown in Figure~\ref{sg8}b. c) Overlay of the GMRT 0.61 GHz contours (in cyan) on the {\it Spitzer} ratio map of 4.5 $\mu$m/3.6 $\mu$m emission (see the dotted-dashed box in Figure~\ref{sg7}a and also Figure~\ref{sg77}).
The ratio map is exposed to a Gaussian smoothing function with a width of 4 pixels. 
In each panel, a star symbol indicates the position of the 6.7 GHz mme. An infrared counterpart (IRc) of the 6.7 GHz mme (i.e. IRcmme) is seen in each color-composite map, 
and is found to be away from the radio continuum emission. In each panel, other symbols are the same as in Figure~\ref{sg1}. Ellipses represent the beam sizes of radio continuum data in the panels.}
\label{sg7}
\end{figure*}

In Figure~\ref{sg7}a, we have overlaid both the GMRT maps on a three 
color-composite map made using the {\it Spitzer} 8.0 $\mu$m (red), 4.5 $\mu$m (green), and 3.6 $\mu$m (blue) images. The composite map shows an extended 4.5 $\mu$m emission associated with 
the HMPO candidate (IRAS 17008-4040~I or IRcmme) without any ionized emission. Previously, an extended green object (EGO) G345.51+0.35 was reported toward the 6.7 GHz mme in the site IRAS 17008-4040 \citep[e.g.,][]{cyganowski08}. 
In general, EGOs associated with the 6.7 GHz masers are speculated to be MYSOs, and are also thought to indicate the existence of the shocked 
gas in molecular outflows \citep[e.g.,][]{cyganowski08}. In both the IRAS sites, the extended 8.0 $\mu$m features associated with the ionized emission are also seen in the map \citep[see also][]{morales09,lopez11}. 
In the composite map, one can also examine the spatial distribution 
of the ionized emission observed in both the GMRT maps. Figure~\ref{sg7}b also displays a three 
color-composite map made using the high-resolution (resolution $\sim$0$''$.8) VVV NIR images (K$_{s}$ (red), H (green), and J $\mu$m (blue)). The color-composite map is also overlaid with the GMRT 0.61 GHz contour. 
The HMPO candidate IRcmme is seen only in the VVV K$_{s}$ image. 
In the VVV HK$_{s}$ images, several embedded point-like sources and noticeable extended emission 
are observed toward both the IRAS sources. 
Figure~\ref{sg7}c shows the {\it Spitzer} ratio map of 4.5 $\mu$m/3.6 $\mu$m emission, tracing the bright emission regions toward both the IRAS sites due to the excess 4.5 $\mu$m emission. 
The process for generating the ratio map can be found in \citet{dewangan16}.  
The {\it Spitzer} 4.5 $\mu$m band is known for hosting a prominent molecular hydrogen line emission ($\nu$ = 0--0 $S$(9); 4.693 $\mu$m), and a hydrogen recombination line Br$\alpha$ (at 4.05 $\mu$m).
However, hydrogen recombination lines (e.g., Br$\alpha$) are generally observed toward the ionized regions. 
In general, such emission shows a very good correlation with the radio continuum emission. However, no radio continuum emission is detected around 
the 6.7 GHz mme. Hence, the absence of the Br$\alpha$ emission 
around the 6.7 GHz mme is expected. 
Thus, the excess 4.5 $\mu$m emission around the 6.7 GHz mme is possibly tracing only the extended molecular hydrogen features, which are generally originated because of an outflow activity. Therefore, it seems that the source IRcmme drives the molecular outflow. 
However, in the direction of both the IRAS sources, the excess emission at 4.5 $\mu$m associated with the ionized clumps may indicate the presence of the Br$\alpha$ features.  

\begin{figure*}
\epsscale{0.8}
\plotone{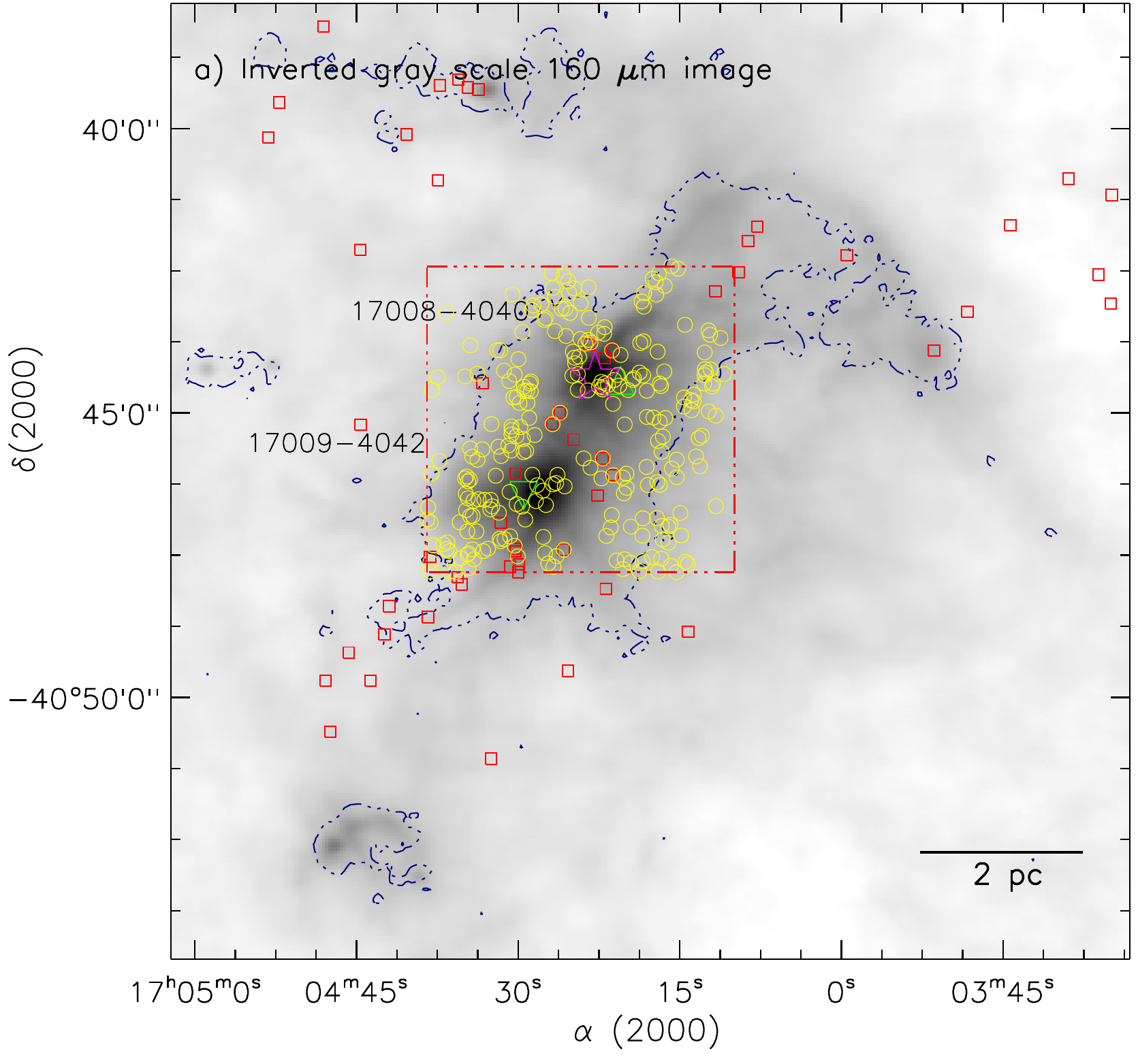}
\epsscale{0.8}
\plotone{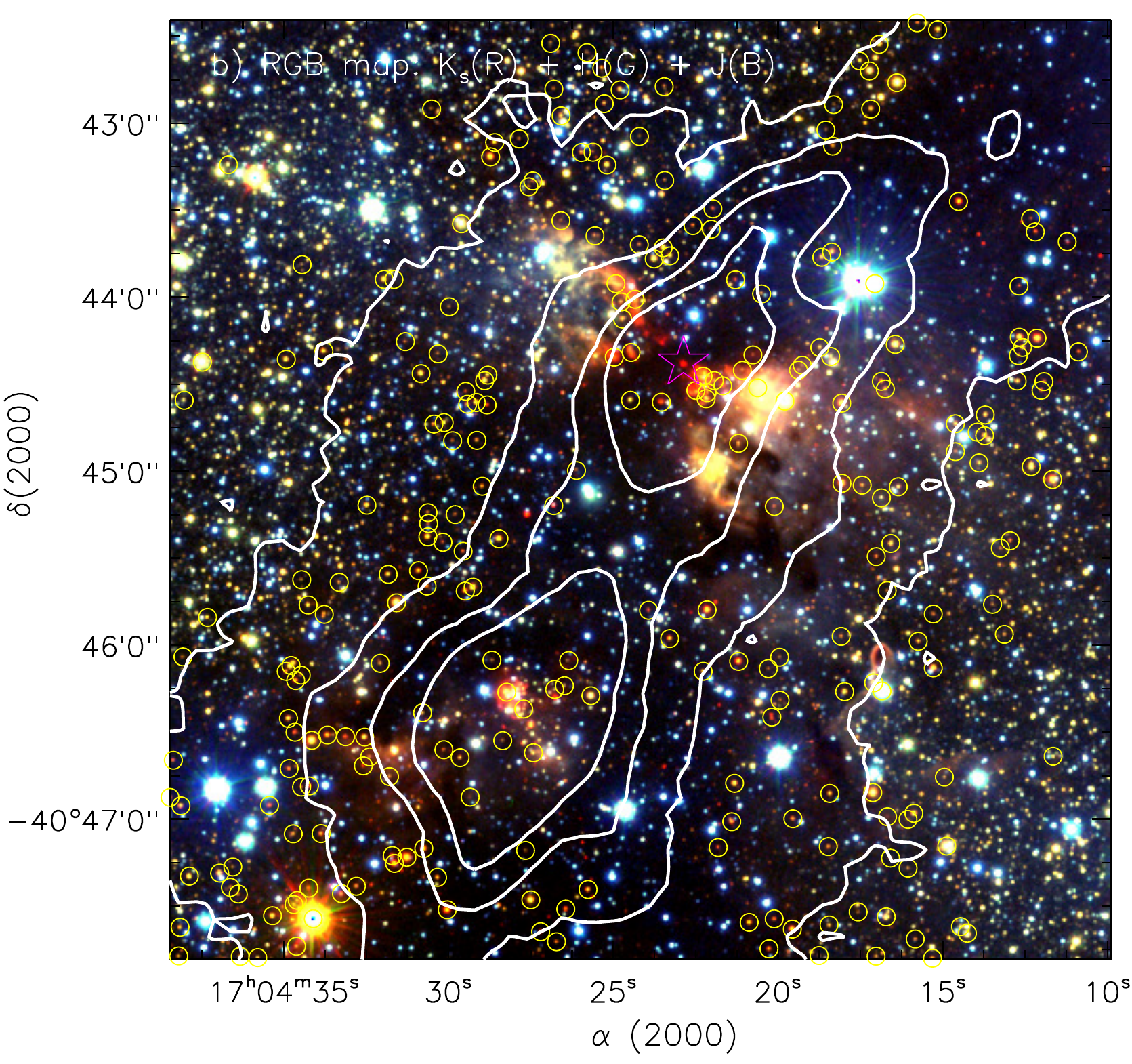}
\caption{a) Overlay of the infrared-excess sources and the ATLASGAL 870 $\mu$m emission contour on an inverted gray scale {\it Herschel} 160 $\mu$m image. 
The ATLASGAL 870 $\mu$m dust continuum contour (in navy) is drawn with a level of 0.13 Jy/beam. 
The YSOs distributed within the molecular cloud boundary are highlighted with open squares (in red) are taken from Figure~\ref{ysg3}b. 
The sources with H$-$K$_{s}$ $>$ 1.8 are shown by yellow circles, and 
are selected only for an area enclosed within a dotted-dashed box (see also Figure~\ref{sg10}b). 
The dotted-dashed box (in red) encompasses the area shown in Figure~\ref{sg10}b.
b) Overlay of the sources with H$-$K$_{s}$ $>$ 1.8 (see yellow circles) and 
the ATLASGAL 870 $\mu$m dust continuum contours on the VVV NIR color-composite map. 
The color-composite map is the same as in Figure~\ref{sg7}b. 
The ATLASGAL 870 $\mu$m contours (in white) are shown with the levels of 0.13, 0.8, 1.8, and 3 Jy/beam. 
In each panel, other symbols are the same as in Figure~\ref{sg1}.}
\label{sg10}
\end{figure*}

In Figure~\ref{sg7}b, we have qualitatively discussed the presence of embedded sources in the VVV NIR color-composite image. 
In order to perform a quantitative analysis, we have selected infrared-excess sources with a 
color (H$-$K$_{s}$) larger than 1.8 mag \citep[or A$_{V}$ = 29 mag;][]{indebetouw05}. 
A majority of these sources do not have J-band 
photometric magnitudes. This analysis is carried out only for the sources located toward both the IRAS sources (see Figure~\ref{sg7}b). We have obtained this color condition through the color-magnitude analysis of a nearby control field (size $\sim$5$'$.4 $\times$ 5$'$.4; central coordinates: $\alpha_{2000}$ = 17$^{h}$04$^{m}$05$^{s}$, $\delta_{2000}$ = $-$41$\degr$04$\arcmin$46$\farcs4$).  
In Figure~\ref{sg10}a, we have presented infrared-excess 
sources (i.e. red squares and yellow circles) overlaid on the inverted gray scale 160 $\mu$m image, which are selected using the {\it Spitzer} 3.6--24 $\mu$m and 
VVV HK$_{s}$ photometric magnitudes. 
The YSOs highlighted with squares are taken from Figure~\ref{ysg3}b, and are distributed within the molecular cloud boundary. 
Circles (in yellow) represent sources with H$-$K$_{s}$ $>$ 1.8 mag, which are selected within a field highlighted by a dotted-dashed box in Figure~\ref{sg10}a. The sources with larger HK$_{s}$ color-excess are also marked on the VVV NIR color-composite map in Figure~\ref{sg10}b. The ATLASGAL 870 $\mu$m emission contours are also overlaid on 
the NIR color-composite map. Our analysis indicates the presence of a group 
of infrared excess sources toward the clumps c1 and c2, 
where the ``hub-filament" configurations are investigated (see Section~\ref{subsec:u2}). 
\subsection{High-resolution NIR images of IRcmme}
\label{subsec:u4}
\begin{figure*}
\epsscale{1.15}
\plotone{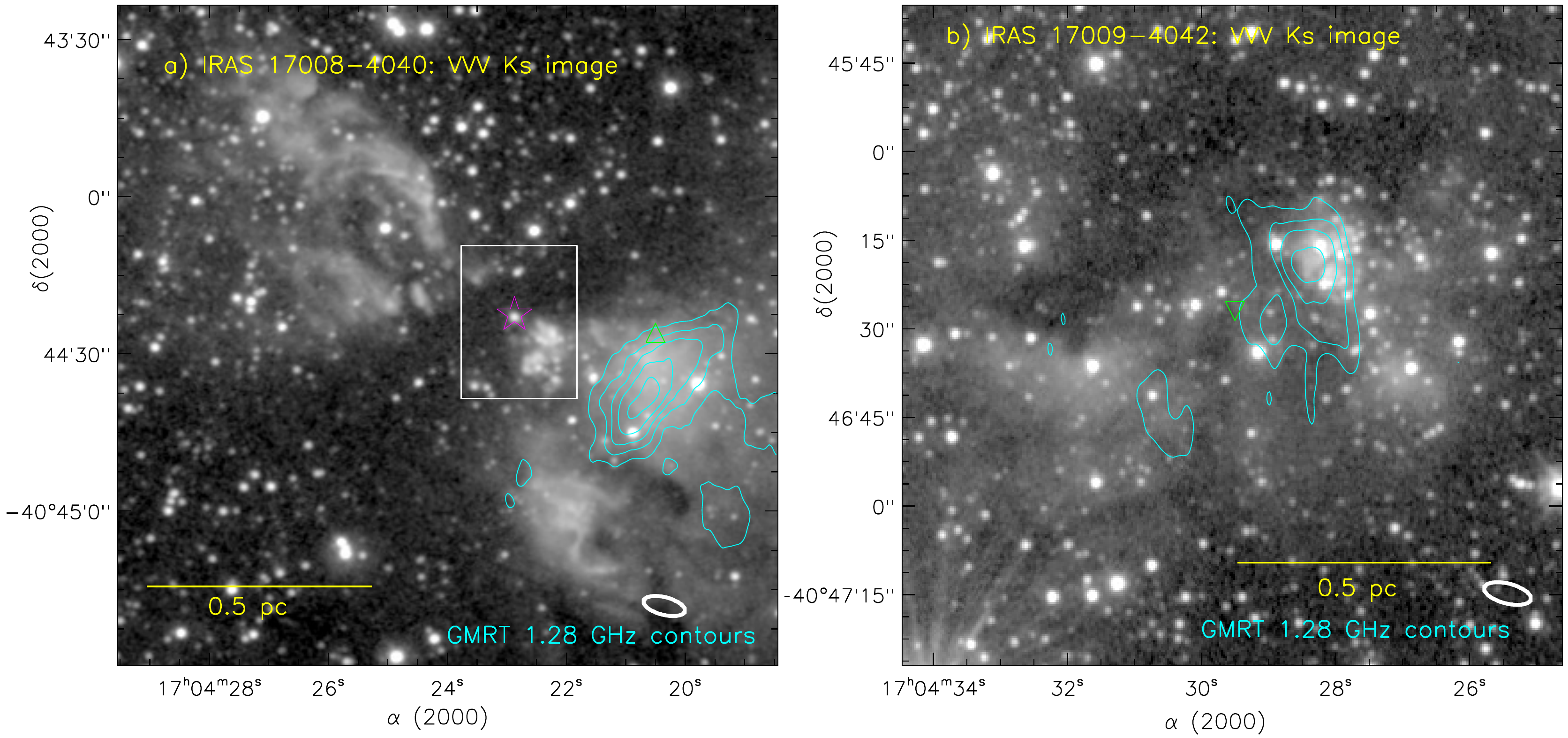}
\epsscale{1.15}
\plotone{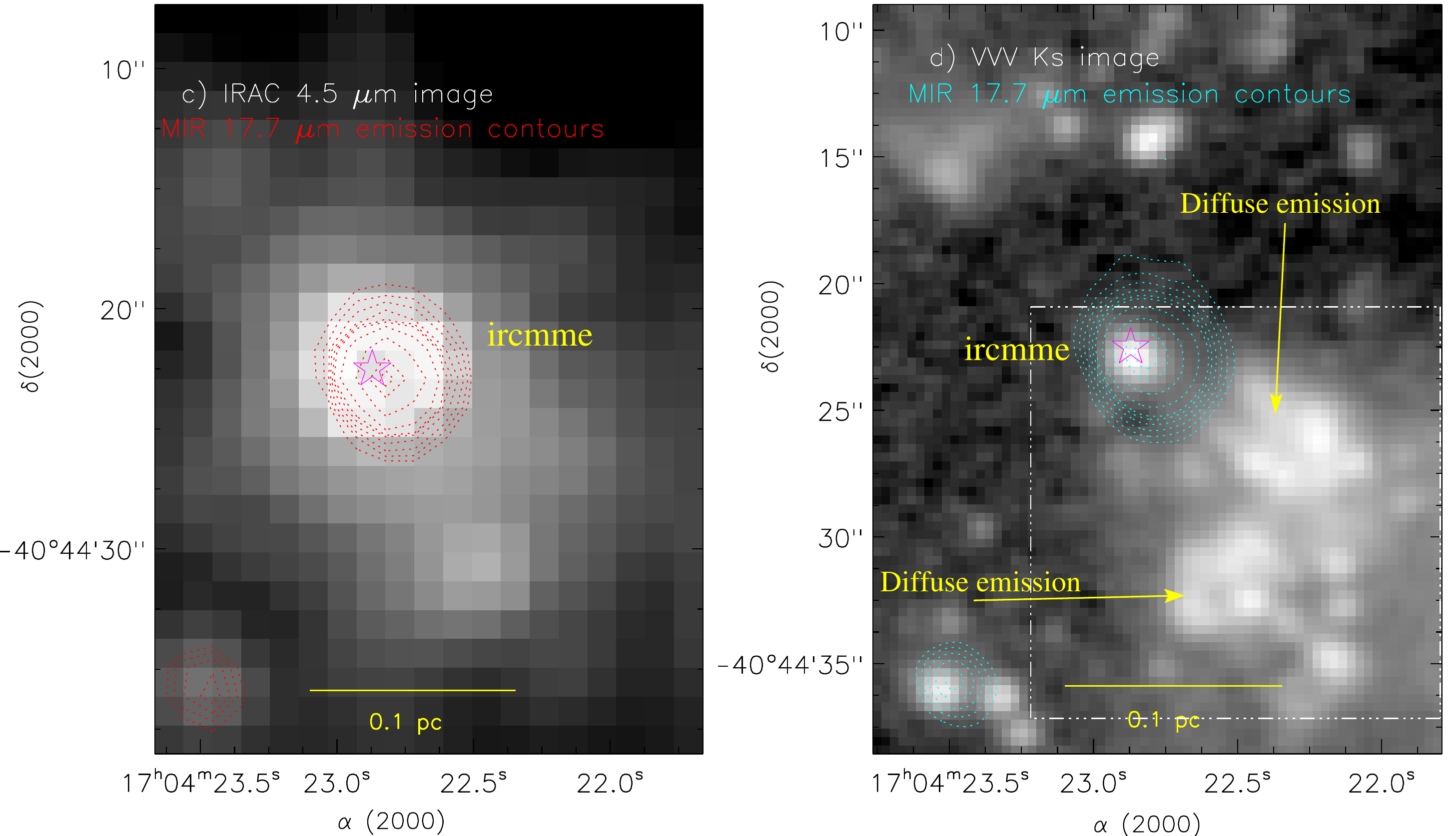}
\caption{a) Overlay of the GMRT 1.28 GHz continuum contours (in cyan) on the 
VVV K$_{s}$ image toward IRAS 17008-4040 (see the dotted-dashed box in Figure~\ref{sg7}b). 
The contour levels are 3.5, 6.5, 9.5, 13, and 17 mJy/beam. 
The solid box (in white) encompasses the area shown in Figures~\ref{sg8}c and~\ref{sg8}d. 
b) Overlay of the GMRT 1.28 GHz continuum contours (in cyan) on the 
VVV K$_{s}$ image toward IRAS 17009-4042 (see the dashed box in Figure~\ref{sg7}b). 
The contour levels are 3.5, 13, 25, and 45 mJy/beam. 
c) Overlay of the TIMMI2 MIR emission contours at 17.7 $\mu$m \citep[in red; resolution $\sim$1$''$;][]{morales09} on 
the {\it Spitzer} 4.5 $\mu$m map (see the solid box in Figure~\ref{sg8}a). 
d) Overlay of the TIMMI2 MIR emission contours at 17.7 $\mu$m (in cyan) on 
the VVV K$_{s}$ image (see the solid box in Figure~\ref{sg8}a). 
The dotted-dashed box (in white) encompasses the area shown in 
Figures~\ref{sg9}a and~\ref{sg9}b. Ellipses (in white) represent the beam sizes of radio continuum data in the panels ``a" and ``b". In the panels ``c" and ``d", the TIMMI2 MIR 17.7 $\mu$m data are taken from \citet{morales09}. In all the panels, other symbols are the same as in Figure~\ref{sg1}.}
\label{sg8}
\end{figure*}

Figures~\ref{sg8}a and~\ref{sg8}b present a zoomed-in view of the sites IRAS 17008-4040 
and IRAS 17009-4042 using the high-resolution VVV K$_{s}$ image, respectively. 
The GMRT 1.28 GHz radio contours are also overlaid on the VVV image. 
Using the TIMMI2 17.7 $\mu$m, {\it Spitzer} 4.5 $\mu$m and VVV K$_{s}$ images, a further zoomed-in view of the HMPO candidate IRcmme is shown in Figures~\ref{sg8}c and~\ref{sg8}d. 
Figures~\ref{sg8}c and~\ref{sg8}d show the {\it Spitzer} 4.5 $\mu$m and VVV K$_{s}$ images overlaid with the TIMMI2 MIR emission contours at 17.7 $\mu$m, respectively. The HMPO candidate IRcmme having a spectral type of O9.5 is found at the peak of the 17.7 $\mu$m emission, as previously reported by \citet{morales09}. 
Noticeable diffuse emission in the VVV K$_{s}$ image is detected in the south-west direction of IRcmme (see arrows in Figure~\ref{sg8}d), which 
is not resolved in the {\it Spitzer} 4.5 $\mu$m image. 
The observed diffuse emission is not associated with any ionized emission. 
It is known that the K$_{s}$ band includes the H$_{2}$ (at 2.12 $\mu$m) and Br$\gamma$ (at 2.16 $\mu$m) lines. 
In the absence of the ionized region, it is unlikely that the diffuse emission 
seen in the K$_{s}$ band is originated because of the NIR hydrogen recombination line like Br$\gamma$. Hence, the diffuse emission could be H$_{2}$ emission excited by the outflow activity. 
In the north-east direction of IRcmme, the {\it Spitzer} 4.5 $\mu$m excess emission as well as the extended emission features 
in the K$_{s}$ image are evident, which are also devoid of the ionized emission. 
Together, we suggest the presence 
of a bipolar outflow (i.e. north-east to south-west direction) associated with IRcmme. 
In general, the existence of the molecular outflow may 
provide an evidence of an accretion process. 

\begin{figure*}
\epsscale{0.89}
\plotone{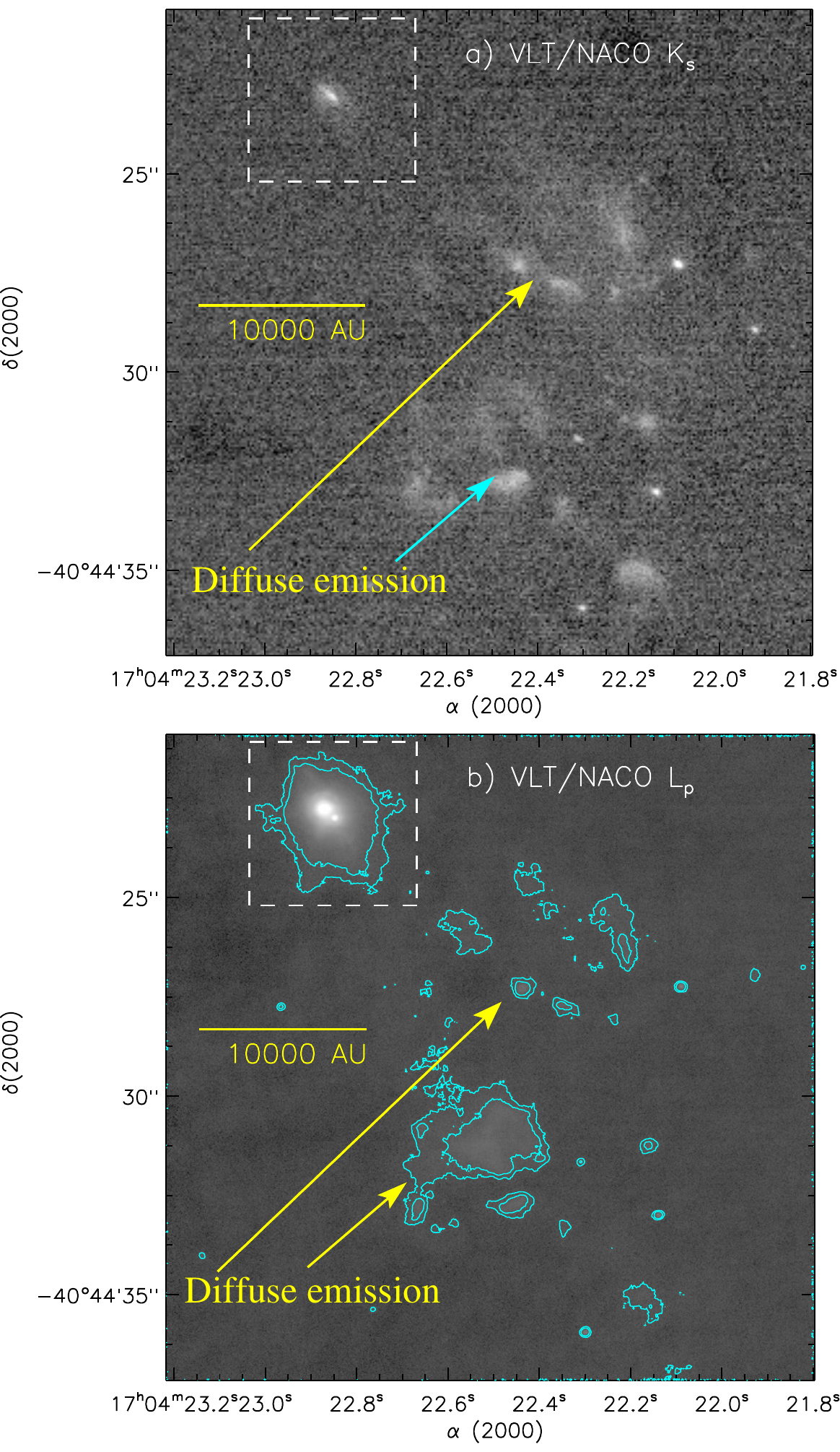}
\caption{The VLT/NACO adaptive-optics K$_{s}$-band (a) and L$^{\prime}$-band (b) 
images of the region toward the source ``IRcmme" (see the dotted-dashed box in Figure~\ref{sg8}d). 
In the panel ``b", L$^{\prime}$ contours (in cyan) are also shown to highlight the faint features.}
\label{sg9}
\end{figure*}

\begin{figure*}
\epsscale{0.87}
\plotone{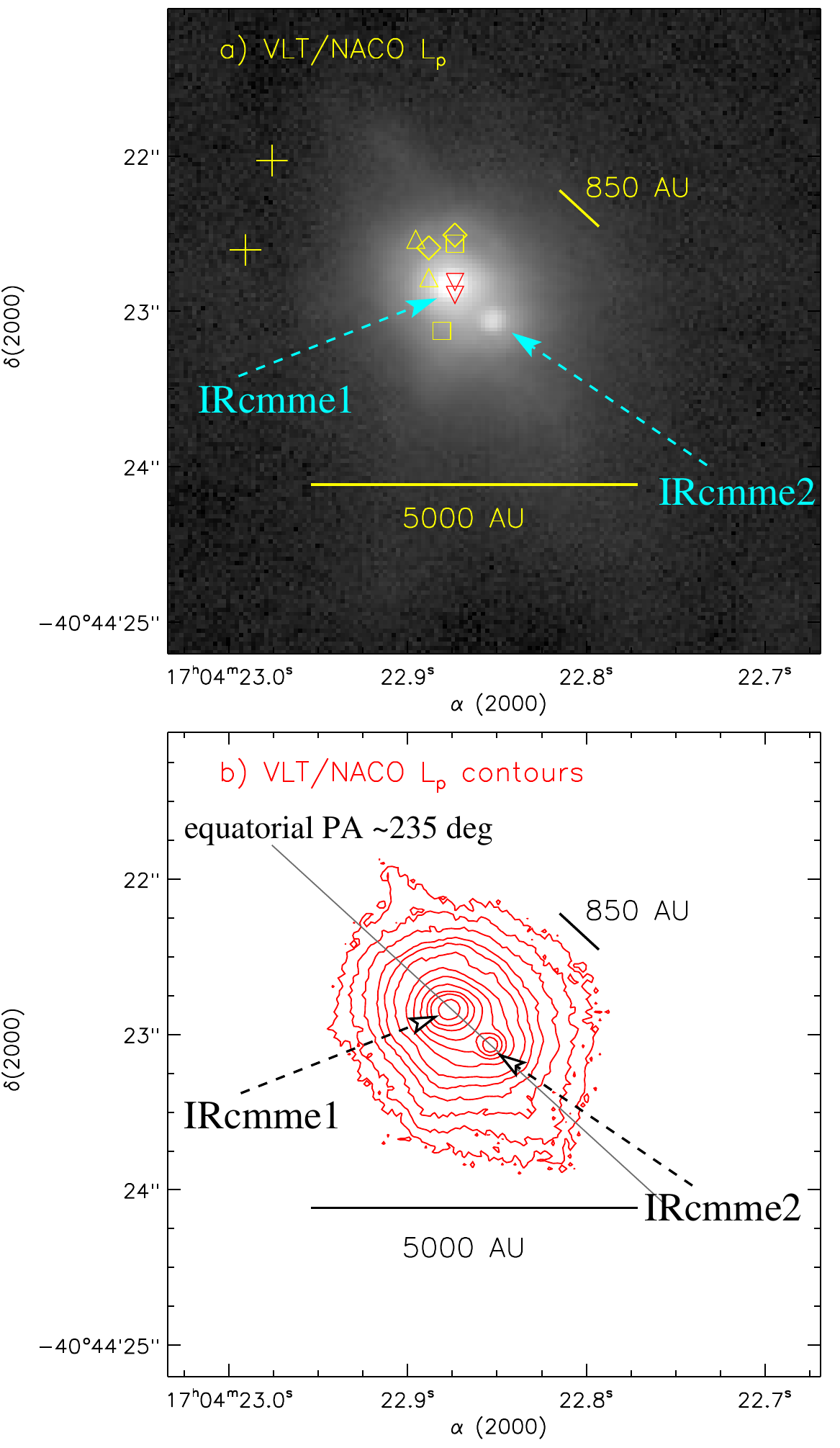}
\caption{a) A zoomed-in view of the source ``IRcmme" using the VLT/NACO L$^{\prime}$ image (see the dashed box in Figures~\ref{sg9}a and~\ref{sg9}b). Ten positions of the 6.7 GHz maser spots \citep[from][]{walsh98} are marked by 
plus symbols (V$_{lsr}$ range = [$-$14, $-$15] km s$^{-1}$), diamonds (V$_{lsr}$ range = [$-$15, $-$17] km s$^{-1}$), triangles (V$_{lsr}$ range = [$-$18, $-$19] km s$^{-1}$), squares (V$_{lsr}$ range = [$-$19, $-$20] km s$^{-1}$), and upside down triangles 
(V$_{lsr}$ range = [$-$20, $-$22.5] km s$^{-1}$). b) A zoomed-in view of the source ``IRcmme" using the VLT/NACO L$^{\prime}$ emission contours (see the dashed box in Figures~\ref{sg9}a and~\ref{sg9}b). 
The NACO L$^{\prime}$ image has resolved the source ``IRcmme" into two point-like sources (i.e. IRcmme1 and IRcmme2), which are embedded in an extended envelope within a scale of 5000 AU. The separation between ``IRcmme1" and ``IRcmme2" is $\sim$900 AU.}
\label{ffsg9}
\end{figure*}

To examine the inner environment of IRcmme, the VLT/NACO adaptive-optics images of IRcmme in K$_{s}$ and L$^{\prime}$ bands are presented in Figures~\ref{sg9}a and~\ref{sg9}b, respectively. 
The VLT/NACO K$_{s}$ image (resolution $\sim$0$''$.2) does not resolve IRcmme. The diffuse emission as seen in the VVV K$_{s}$ image is observed in both the VLT/NACO K$_{s}$, and L$^{\prime}$ images. 
The VLT/NACO L$^{\prime}$ image (resolution $\sim$0$''$.1) resolves the HMPO candidate IRcmme into two point-like sources (see Figure~\ref{sg9}b). 
Using the VLT/NACO gray scale L$^{\prime}$ image, Figure~\ref{ffsg9}a displays a zoomed-in view of the IRcmme. 
The positions of the 6.7 GHz maser spots \citep[from][]{walsh98} are also marked in the L$^{\prime}$ image (see Figure~\ref{ffsg9}a). 
In Figure~\ref{ffsg9}b, we present the contours of L$^{\prime}$ image for more clarity. 
In Figures~\ref{ffsg9}a and~\ref{ffsg9}b, we have designated the resolved two sources as IRcmme1 ($\alpha_{2000}$ = 17$^{h}$04$^{m}$22$^{s}$.88; 
$\delta_{2000}$ = $-$40$\degr$44$\arcmin$22$\farcs86$) and IRcmme2 ($\alpha_{2000}$ = 17$^{h}$04$^{m}$22$^{s}$.85; $\delta_{2000}$ = $-$40$\degr$44$\arcmin$23$\farcs08$). 
The spatial separation between these two sources is about 850 AU. 
An extended envelope-like feature is observed within a scale of 5000 AU in the north-east and south-west direction, and these two resolved sources are embedded within this envelope. 
The spatial orientation of the envelope appears to be aligned with the large-scale outflow features as discussed earlier in this section. 
The source IRcmme1 is spatially seen more extended, while IRcmme2 looks like a point-like source. 
We have found that the positions of the 6.7 GHz maser spots correlate more with the source IRcmme1. The 6.7 GHz mme is believed to be turned on after the onset of the outflow \citep[e.g.,][]{devilliers15}. 
Ten positions of the 6.7 GHz maser spots are shown by different symbols, and are detected with a large V$_{lsr}$ spread (see plus symbols (V$_{lsr}$ range = [$-$14, $-$15] km s$^{-1}$), diamonds (V$_{lsr}$ range = [$-$15, $-$17] km s$^{-1}$), triangles (V$_{lsr}$ range = [$-$18, $-$19] km s$^{-1}$), squares (V$_{lsr}$ range = [$-$19, $-$20] km s$^{-1}$), and upside down triangles 
(V$_{lsr}$ range = [$-$20, $-$22.5] km s$^{-1}$) in Figure~\ref{ffsg9}a). These results hint the outflow activity associated 
with the source IRcmme1. 
Hence, the source IRcmme1 may be considered as the main massive protostar/HMPO that appears to drive the molecular outflow. 
It seems that the HMPO candidate IRcmme1 is still in accretion phase and has not yet excited an UCH\,{\sc ii} region. 
As we know from the study of the formation of low-mass stars, the accretion disk is 
a natural outcome of the accretion process \citep[e.g.,][]{takakuwa17}. Hence, with the resolution of 0$''$.1 (or 240 AU for a distance 2.4 kpc), the VLT/NACO L$^{\prime}$ image 
is unable to resolve the disk associated with the source IRcmme1. 
\section{Discussion}
\label{sec:disc}
A proper understanding of the ongoing physical mechanism in a given star-forming region not only requires a morphological overview of a large area surrounding the target region, it also needs high-resolution observations for obtaining the finer details of the individual star-forming cores/clumps. 
In this context, the sites IRAS 17008-4040 and IRAS 17009-4042 have been explored using the observational data sets with different resolutions. 
\subsection{Star formation scenario in IRAS 17008-4040 and IRAS 17009-4042}
\label{subsec:u5}
We have carried out a careful analysis of the multi-wavelength data of IRAS 17008-4040 and IRAS 17009-4042.
The massive ATLASGAL clump c1 (M$_{clump}$ $\sim$2430 M$_{\odot}$ and R$_{c}$ $\sim$3 pc) hosting the site IRAS 17008-4040 is evident 
with the ongoing MSF activities, and contains a group of infrared-excess sources/YSOs. 
On the other hand, the site IRAS 17009-4042 is embedded within another massive ATLASGAL clump c2 
(M$_{clump}$ $\sim$2900 M$_{\odot}$ and R$_{c}$ $\sim$2.15 pc), where a cluster of infrared-excess sources and several B-type stars are observationally found. The presence of several B-type stars is inferred from the radio continuum observations. 
Hence, the intense star formation activities are depicted toward both the clumps. 
These clumps also have relatively higher dust temperatures compared to other regions in the molecular cloud (see Figure~\ref{ysg2}b). 
In other words,  the star formation activities are more concentrated toward these two massive clumps in the entire molecular cloud. 

In the ATLASGAL 870 $\mu$m image, these two clumps, containing massive stars and cluster of infrared-excess sources, appear like fragments in the elongated filamentary feature (see Figure~\ref{sg5}c). 
Each of these massive clumps is investigated as the junction of at least three parsec-scale filaments in the {\it Herschel} 160 $\mu$m image.  
Such configuration is known as a ``hub-filament" system \citep[e.g.,][]{myers09}, and in the literature, many such sites have been reported \citep[e.g.,][]{myers09,schneider12,hennemann12,liu12,liu16,peretto13,dewangan15b,baug15,baug18,dewangan17,yuan18,williams18}. 
It has been thought for the ``hub-filament" configuration that the gas can be funneled toward the junction/hub through the parsec-scale filaments, where very intense star formation activities are observed \citep[e.g.,][]{kirk13,liu16,baug18}. Dust temperature is also expected to be higher toward the ``hub/junction" due to the star formation activities, and has been also observed in our selected target sources. Due to the coarser beam size of the ThrUMMS CO line data, we are 
unable to provide the kinematical insights into the proposed scenario in both the IRAS sites.
However, our results are suggestive and promising to explain the existence of massive clumps with the ongoing star formation activities.
Hence, high-resolution molecular line observations with ALMA can provide more detailed spatial and velocity structures 
in the ``hub-filament" systems, which will help us to further understand the ongoing physical process.
\subsection{Inner circumstellar environment of the youngest HMPO candidate}
\label{subsec:u6}
In the literature, IRAS 17008-4040~I has been suggested as a very promising HMPO candidate with a spectral type of O9.5 \citep[e.g.,][]{morales09}. 
In this paper, we have referred to this HMPO candidate as an IRc of the 6.7 GHz mme (i.e. IRcmme) that is located at the peak of the 870 $\mu$m emission. 
No radio continuum emission has been detected towards the HMPO candidate that also appears to drive a large-scale molecular outflow in the north-east to south-west direction. 
As mentioned earlier, the detection of the 6.7 GHz mme indicates the early phases of MSF ($<$ 0.1 Myr). 
Hence, the source IRcmme can be considered as a rare massive young stellar object (MYSO), like W42-MME as reported by \citet{dewangan15}. 
As previously highlighted, in the direction of the source G345.50+0.35/IRAS 17008-4040~I/IRcmme, two molecular cores (i.e. G345.50M and G345.50S) have been reported using the high-resolution ALMA data with a resolution of $\sim$0$''$.2 \citep[see Figure~2 in][]{cesaroni17}. The complex morphology of the SiO(5--4) line (spectral window: 216.976--218.849 GHz) emission was also reported toward the core G345.50M \citep[see Figure~15 in][]{cesaroni17}. 
However, there is no satisfactory explanation available for the existence of the extended and complex SiO(5--4) emission toward the core G345.50M.

Within a scale of 900 AU, the VLT/NACO L$^{\prime}$ image has resolved the single object IRcmme into two sources (i.e. IRcmme1 and IRcmme2). 
The spatial distribution of these two sources seems parallel to a line having an equatorial position angle (EPA) of 235$\degr$ 
(see a solid gray line in Figure~\ref{ffsg9}b). 
In the VLT/NACO L$^{\prime}$ image, these two sources are also spatially located inside the envelope-like feature, which is extended within 
a scale of 5000 AU in the north-east and south-west direction. We have compared the VLT/NACO L$^{\prime}$ image with the published ALMA molecular maps by \citet{cesaroni17}. 
Molecular emission traced in the ALMA molecular maps is spatially seen toward the extended NIR envelope feature \citep[see Figure~11 in][]{cesaroni17}, confirming the existence of the envelope-like feature in IRcmme. 
The molecular envelope feature hosts the ALMA core G345.50M, which shows an irregular shape. 
There is no infrared emission observed toward the other ALMA core G345.50S in the VLT/NACO L$^{\prime}$ image. 
Furthermore, we find that the sources IRcmme1 and IRcmme2 are seen toward the ALMA core G345.50M. 
It implies that these two sources could be responsible for the observed SiO(5--4) emission toward the core G345.50M. 
The SiO emission was studied in the velocity ranges of [$-$23.8, $-$18.4] and [$-$15.7, $-$10.3] km s$^{-1}$. 
Hence, it is possible that both these sources drive the molecular outflows (see diffuse emission in Figures~\ref{sg9}a and~\ref{sg9}b). 
However, several 6.7 GHz maser spots observed by \citet{walsh98} are exclusively seen in the direction of the source IRcmme1 that spatially appears more extended compared to the point-like source IRcmme2. 
An observed velocity spread in the 6.7 GHz maser spots also favours the outflow activity associated with IRcmme1. 
Hence, we find the source IRcmme1 as the main massive protostar/HMPO that is going through the accretion phase. 
It is also possible that the source IRcmme1 might be associated with 
its circumstellar disk below 100 AU, which is not spatially resolved by the NACO and ALMA data with the resolutions of 0$''$.1--0$''$.2.

During the early phases of the evolution of massive stars, a large mass reservoir is expected to be available \citep[e.g.,][]{tobin16}. Hence, massive stars have tendency to produce binaries (or multiple systems) in the early phases 
of their evolution \citep[e.g.,][]{krumholz07,kratter08,tobin16}. 
\citet{tobin16} discussed several possible mechanisms 
(i.e. the turbulent fragmentation of the molecular cloud;  
the thermal fragmentation of strongly perturbed, rotating, and infalling core; and/or the 
fragmentation of a gravitationally unstable circumstellar disk) to explain the observed multiple systems.
They also argued that the knowledge of the companion separations in multiple systems helps to understand the ongoing physical process. Considering the separation between IRcmme1 and IRcmme2 (i.e. $\sim$850 AU), 
the core fragmentation process is likely mechanism for these sources in the IRcmme system. As highlighted earlier, these sources are embedded within the single and rotating ALMA core. 
Hence, our interpretation is also supported by the ALMA data. 
\section{Summary and Conclusions}
\label{sec:conc}
In this paper, we have studied the inner and large scale physical environments of IRAS 17008-4040 and IRAS 17009-4042 using a multi-scale and multi-wavelength approach. 
The major results of this work are presented below.\\
$\bullet$ The molecular cloud associated with the sites IRAS 17008-4040 and IRAS 17009-4042 
is studied in a velocity range of [$-$22, $-$10] km s$^{-1}$. 
The observed sub-mm emission in the ATLASGAL 870 $\mu$m continuum map traces the denser parts in the molecular cloud, where nine 
clumps (M$_{clump}$ $\sim$35--2900 M$_{\odot}$) are detected.\\
$\bullet$ An extended and elongated morphology is also observed in the ATLASGAL 870 $\mu$m continuum map, and contains two massive clumps c1 and c2.\\ 
$\bullet$ The site IRAS 17008-4040 is embedded within the clump c1 (M$_{clump}$ $\sim$2430 M$_{\odot}$ and R$_{c}$ $\sim$3 pc), while the clump c2 (M$_{clump}$ $\sim$2900 M$_{\odot}$ and R$_{c}$ $\sim$2.15 pc) hosts the site IRAS 17009-4042.\\
$\bullet$ The clumps c1 and c2 are seen at the junction of multiple {\it Herschel} filaments (i.e. ``hub-filament" systems). 
In these systems, several parsec-scale embedded filaments 
are identified at 160 $\mu$m, and they are radially pointed toward the massive clumps (at T$_{d}$ $\sim$25--32 K).\\
$\bullet$ With the analysis of the {\it Spitzer} and VVV photometric data, a cluster of infrared-excess sources is depicted toward the clumps c1 and c2, suggesting the star formation activities in both the clumps.\\ 
$\bullet$ High-resolution GMRT radio continuum maps at 0.61 GHz (beam size $\sim$10$''$.1 $\times$ 4$''$.6)
and 1.28 GHz (beam size $\sim$5$''$.3 $\times$ 1$''$.7) have detected the H\,{\sc ii} regions toward the clumps c1 and c2, and each of the H\,{\sc ii} regions is excited by at least a B-type star.\\  
$\bullet$ In the site IRAS 17009-4042, a single ATCA radio peak at 2.5 GHz is resolved into two radio sources 
in the 1.28 GHz map (see ionized clumps n3 and n4 in Table~\ref{stab1}; spectral types = B0.5V), where the infrared-excess sources are distributed. \\   
$\bullet$ The radio clump associated with IRAS 17009-4042 is found to be thermal in nature. However, a flat spectral index is obtained for the radio clump associated with IRAS 17008-4040, implying the presence of non-thermal and thermal emission in the radio clump.\\
$\bullet$ Radio continuum emission is not found toward a previously known IRc of the 6.7 GHz mme (i.e. IRcmme). The source IRcmme has been characterized as a genuine massive protostar candidate (with a spectral type of O9.5) in a very early evolutionary stage, before the onset of an UCH\,{\sc ii} phase.\\ 
$\bullet$ In the site IRAS 17008-4040, at least two B-type sources and the HMPO candidate IRcmme 
(without any radio emission) are investigated, illustrating the presence of different early evolutionary stages of MSF. \\
$\bullet$ The inner circumstellar environment of IRcmme is examined using the VLT/NACO adaptive-optics 
L$^{\prime}$ observations (resolution $\sim$0\farcs1). The HMPO candidate IRcmme is resolved into two sources (i.e. IRcmme1 and IRcmme2) in the inner 900~AU, and one of them (i.e. IRcmme1) is associated with several 6.7 GHz maser spots.\\ 
$\bullet$ The sources, IRcmme1 and IRcmme2, are found to be embedded in the ALMA core G345.50M, which is also located within the extended circumstellar envelope in a scale of 5000 AU.\\ 
$\bullet$ The detection of two NACO sources (i.e. IRcmme1 and IRcmme2) in the ALMA core G345.50M explains 
the complex morphology of the observed ALMA SiO(5--4) emission.\\ 
$\bullet$ In the light of the published ALMA results, the core fragmentation process appears to be responsible for the observed separation between IRcmme1 and IRcmme2 (i.e. $\sim$850 AU).\\

Together, in each IRAS site, the junction of the filaments (i.e. massive clump) is identified 
with the ongoing MSF activities and the cluster of infrared-excess sources. These observational 
outcomes are in agreement with the ``hub-filament" systems as proposed by \citet{myers09}. 
\acknowledgments 
We thank the anonymous reviewers for several useful comments. 
The research work at the Physical Research Laboratory is funded by the Department of Space, Government of India.
This publication made use of data products from the Two Micron All Sky 
Survey (a joint project of the University of Massachusetts and 
the Infrared Processing and Analysis Center / California Institute of Technology, funded by NASA and NSF), archival 
data obtained with the {\it Spitzer} Space Telescope (operated by the Jet Propulsion Laboratory, California Institute 
of Technology under a contract with NASA). 
TB acknowledges funding from the National Natural Science Foundation of China through grant 11633005.
%

%
\begin{deluxetable}{cccccccccccccccc}
\tablewidth{0pt} 
\tablecaption{ATLASGAL 870 $\mu$m dust continuum clumps from \citet{urquhart18} in our selected target field (see Figure~\ref{sg1}a). We have listed ID, 
equatorial coordinates, 870 $\mu$m peak flux density (P$_{870}$), 
870 $\mu$m integrated flux density (S$_{870}$), radial velocity (V$_{lsr}$), distance, clump effective radius ($R_\mathrm{c}$), dust temperature ($T_\mathrm{d}$), and clump mass ($M_\mathrm{clump}$). The positions of IRAS 17008-4040 and IRAS 17009-4042 are embedded in the clumps c1 and c2, respectively. In Figures~\ref{sg1}a and~\ref{sg1}b, three clumps (i.e. c10, c11, and c12) are highlighted by diamonds, while nine clumps (i.e. c1--c9) are shown by circles. \label{tab1}} 
\tablehead{ \colhead{ID} & \colhead{RA} & \colhead{DEC}& \colhead{P$_{870}$}& \colhead{S$_{870}$} & \colhead{V$_{lsr}$}&\colhead{distance}& \colhead{R$_\mathrm{c}$}& \colhead{T$_{d}$}& \colhead{log$M_\mathrm{clump}$}\\
\colhead{} &  \colhead{(J2000)} & \colhead{(J2000)} & \colhead{(Jy/beam)}& \colhead{(Jy)}& \colhead{(km s$^{-1}$)} & \colhead{(kpc)}& \colhead{(pc)}& \colhead{(K)}&\colhead{($M_\odot$)}}
\startdata 

c1  & 17:04:23.10 & -40:44:26.72  & 11.69 & 133.92 & -17.0  &  2.4   & 3.04  & 30.0 &  3.386 \\
c2  & 17:04:28.33 & -40:46:24.61  & 17.27 & 143.50 & -17.3  &  2.4   & 2.15  & 27.3 &  3.463 \\
c3  & 17:03:52.34 & -40:43:45.94  &  0.46 &   7.72 & -16.9  &  2.4   & 0.64  & 21.9 &  2.325 \\
c4  & 17:04:04.03 & -40:42:03.59  &  0.62 &  14.43 & -15.8  &  2.4   & 1.14  & 16.5 &  2.781 \\
c5  & 17:04:27.33 & -40:39:14.66  &  0.39 &   3.44 & -17.4  &  2.4   & 0.28  & 16.5 &  2.159 \\
c6  & 17:04:33.13 & -40:39:28.90  &  0.52 &   4.76 & -16.7  &  2.4   & 0.28  & 25.2 &  2.031 \\
c7  & 17:04:52.85 & -40:38:38.11  &  0.40 &   0.71 & -17.3  &  2.4   & 0.28  & 14.2 &  1.580 \\
c8  & 17:05:03.72 & -40:37:06.83  &  0.36 &   1.99 & -17.6  &  2.4   & 0.28  & 13.5 &  2.065 \\
c9  & 17:05:09.54 & -40:35:09.92  &  0.45 &   0.92 & -16.3  &  2.4   & 0.28  & 13.8 &  1.713 \\
c10 & 17:04:45.40 & -40:52:16.13  &  0.64 &   3.72 &  -4.6  &  2.4   & 0.47  & 23.9 &  1.955 \\
c11 & 17:04:59.09 & -40:44:12.72  &  0.37 &   2.04 & -26.2  &  2.4   & 0.28  & 14.2 &  2.039 \\
c12 & 17:03:21.40 & -40:55:32.00  &  0.66 &   4.47 & -23.8  &  2.4   & 0.92  & 13.4 &  2.422 \\
\enddata  
\end{deluxetable}

\begin{table*}
\tiny
\setlength{\tabcolsep}{0.15in}
\centering
\caption{List of several surveys used in this paper.}
\label{ftab1}
\begin{tabular}{lcccccccr}
\hline 
  Survey  &  Wavelength/Frequency       &  Resolution ($\arcsec$)        &  Reference  \\   
\hline
\hline 
Giant Metre-wave Radio Telescope (GMRT) archival data                               & 0.61, 1.28 GHz                           & $<$10                 & Proposal-ID: 11SKG01\\
Three-mm Ultimate Mopra Milky Way Survey (ThrUMMS)                                  & 115.27, 110.2 GHz & $\sim$72              & \citet{barnes15}\\
APEX Telescope Large Area Survey of the Galaxy (ATLASGAL)                           & 870 $\mu$m                                  & $\sim$19.2            & \citet{schuller09}\\
{\it Herschel} Infrared Galactic Plane Survey (Hi-GAL)                              & 70, 160, 250, 350, 500 $\mu$m               & $\sim$5.8--37         & \citet{molinari10}\\
{\it Spitzer} Galactic Legacy Infrared Mid-Plane Survey Extraordinaire (GLIMPSE) & 3.6, 4.5, 5.8, 8.0 $\mu$m                   & $\sim$2               & \citet{benjamin03}\\
ESO 8.2m Very Large Telescope (VLT) adaptive-optics near-infrared archival data  & 2.18, 3.8 $\mu$m                            & $\sim$0.2, $\sim$0.1  & Proposal-ID: 083.C-0582(A) \\ 
Vista Variables in the V\'{\i}a L\'{\i}actea (VVV)                                  & 1.25--2.2 $\mu$m                            & $\sim$0.8             & \citet{minniti10}\\ 
Two Micron All Sky Survey (2MASS)                                                   & 1.25--2.2 $\mu$m                            & $\sim$2.5             & \citet{skrutskie06}\\
\hline          
\end{tabular}
\end{table*}

\begin{table*}
\setlength{\tabcolsep}{0.15in}
\centering
\caption{Physical properties of the ionized clumps observed in the GMRT 0.61 and 1.28 GHz radio continuum maps. 
Table provides ID, 
equatorial coordinates, deconvolved effective radius of the ionized clump ($R_\mathrm{HII}$), total flux (S${_\nu}$), 
Lyman continuum photons (log$N_\mathrm{uv}$), and radio spectral type. 
Eight ionized clumps (s1--s8) are identified in the GMRT 0.61 GHz radio map (see Figure~\ref{sg77}c), while 
five radio sources (n1--n5) are traced in the GMRT 1.28 GHz radio map (see Figure~\ref{sg77}d).} 
\label{stab1}
\begin{tabular}{lccccccccr}
\hline 
  ID  &  RA     &  Dec    &  $R_\mathrm{HII}$   & S${_\nu}$  & log$N_\mathrm{uv}$    &Spectral Type &Frequency \\  
      & (J2000) & (J2000) &  (pc)               &  (Jy)      &  (s$^{-1}$)           &              & (GHz)\\  
\hline
\hline 
  s1 & 17:04:20.5 &  $-$40:44:20.5 &  0.24 & 0.220 & 46.96 &  B0V--O9.5V &  0.61   \\ 
  s2 & 17:04:20.1 &  $-$40:44:45.7 &  0.25 & 0.170 & 46.85 &  B0V--O9.5V &  0.61   \\ 
  s3 & 17:04:28.3 &  $-$40:46:08.5 &  0.18 & 0.138 & 46.76 &  B0V--O9.5V &  0.61   \\	
  s4 & 17:04:27.6 &  $-$40:46:36.1 &  0.10 & 0.011 & 45.68 &  B1V--B0.5V &  0.61   \\ 
  s5 & 17:04:28.7 &  $-$40:46:40.9 &  0.14 & 0.036 & 46.17 &  B1V--B0.5V &  0.61   \\ 
  s6 & 17:04:29.8 &  $-$40:46:08.5 &  0.09 & 0.010 & 45.60 &  B1V--B0.5V &  0.61   \\ 
  s7 & 17:04:32.0 &  $-$40:46:14.5 &  0.15 & 0.024 & 45.99 &  B1V--B0.5V &  0.61   \\ 
  s8 & 17:04:31.0 &  $-$40:46:49.3 &  0.14 & 0.019 & 45.91 &  B1V--B0.5V &  0.61   \\ 
  n1 & 17:04:20.7 &  $-$40:44:38.5 &  0.16 & 0.482 & 47.33 &  B0.5V--B0V &  1.28  \\	   
  n2 & 17:04:19.4 &  $-$40:45:00.1 &  0.06 & 0.039 & 46.25 &  B1V--B0.5V &  1.28  \\	   
  n3 & 17:04:28.5 &  $-$40:46:20.5 &  0.12 & 0.638 & 47.46 &  B0.5V--B0V &  1.28  \\	   
  n4 & 17:04:29.0 &  $-$40:46:30.1 &  0.08 & 0.140 & 46.80 &  B0.5V--B0V &  1.28  \\ 
  n5 & 17:04:30.4 &  $-$40:46:49.3 &  0.06 & 0.039 & 46.25 &  B1V--B0.5V &  1.28  \\ 

\hline          
\end{tabular}
\end{table*}

\end{document}